\documentclass[12pt]{article}
\usepackage{cite}
\usepackage{amsmath}
\allowdisplaybreaks
\usepackage{arydshln}
\usepackage{multirow}
\usepackage{hyperref}
\usepackage{graphicx}
\graphicspath{{./}{./figures/}}

\newcommand{\vep}{\varepsilon}
\newcommand{\ep}{\epsilon}
\newcommand{\MSbar}{\overline{\mbox{MS}}}

\newcommand{\os}{\mbox{\scriptsize OS}}
\newcommand{\z}[1]{\zeta_{#1}}
\newcommand{\A}[1]{a_{#1}}
\newcommand{\M}[1]{M_{#1}}

\newcommand{\Dalpha}{\Delta\alpha}
\newcommand{\aemnfpi}{\left({\bar{\alpha}^{(n_f)}(\mu)\over\pi}\right)}
\newcommand{\aembarpi}{\left(\frac{\bar{\alpha}}{\pi}\right)}
\newcommand{\aemOSvpi}{\frac{\alpha}{4\*\pi}}
\newcommand{\aemOSpi}{\frac{\alpha}{\pi}}
\newcommand{\Log}[1]{\log\left(#1\right)}
\newcommand{\Logn}[2]{\log^{#2}\!\left(#1\right)}
\newcommand{\LmudMs}[1]{\log\left(\frac{\mu^2}{M_{#1}^2}\right)}

\newcommand{\Lmudmhs}{\log\left({\mu^2\over \overline{m}_h^2}\right)}

\newcommand{\LL}[1]{L_{qM_{#1}}}
\newcommand{\LLn}[2]{L^{#2}_{qM_{#1}}}
\newcommand{\LLmu}[1]{L_{\mu M_{#1}}}
\newcommand{\LLmun}[2]{L^{#2}_{\mu M_{#1}}}
\newcommand{\LLmumh}{L_{\mu\overline{m}_{h}}}
\newcommand{\LLmumhn}[1]{L^{#1}_{\mu\overline{m}_{h}}}
\newcommand{\nh}{n_{h}}
\newcommand{\nl}{n_{l}}
\newcommand{\Me}{M_e}
\newcommand{\Mm}{M_{\mu}}
\newcommand{\Mt}{M_{\tau}}
\newcommand{\p}{\phantom{-}}

\begin{document}    

\begin{titlepage}
\noindent
\mbox{}\hfill{\bf{\mbox{}}}
%

\vspace{0.5cm}
\begin{center}
  \begin{Large}
    \begin{bf}
      \begin{center}
      Leptonic contributions to the effective electromagnetic
      coupling at four-loop order in QED
      \end{center}
    \end{bf}
  \end{Large}
  \vspace{0.8cm}

  \begin{large}
  \end{large}
  \vskip .7cm
{\small {\em  Christian Sturm%
\footnote{Email: \href{mailto:Christian.Sturm@physik.uni-wuerzburg.de}{Christian.Sturm@physik.uni-wuerzburg.de}} \\
Universit{\"a}t W{\"u}rzburg,\\
Institut f{\"u}r Theoretische Physik und Astrophysik,\\
Emil-Hilb-Weg 22,\\
D-97074 W{\"u}rzburg, \\
Germany}}\\[2.5cm]
\vspace{2cm}
{\bf Abstract}
\end{center}
\begin{quotation}
  \noindent
  The running of the effective electromagnetic coupling is for many
  electroweak observables the dominant correction. It plays an important
  role for deriving constraints on the Standard Model in the context of
  electroweak precision measurements.  We compute the four-loop QED
  corrections to the running of the effective electromagnetic coupling
  and perform a numerical evaluation of the different gauge invariant
  subsets.
\end{quotation}
\end{titlepage}
\mbox{}\vspace*{0.6cm}
\section{Introduction}
\label{sec:Introduction}
The input parameters of Quantum electrodynamics (QED), like the
fine-structure constant or the masses of the charged leptons are known
with high precision. For example, the fine-structure constant has been
measured at the level of $\sim0.32$~ppb~\cite{Mohr:2012tt} or the mass
of the electron in units of MeV is known at the level of
$\sim22$~ppb~\cite{Mohr:2012tt}. Also from theory side physical
quantities in QED can be computed to high accuracy in perturbation
theory, since the electromagnetic coupling constant is a small
parameter.

QED corrections play an important role for electroweak precision
observables where the effective electromagnetic coupling at the
$Z$-boson mass scale enters. The change from the fine-structure constant
$\alpha$, defined at zero momentum transfer, to the effective
electromagnetic coupling at the $Z$-boson mass scale is mediated by the
photon vacuum polarization function.  At the $Z$-boson mass scale the
effective electromagnetic coupling is governed by the parameter
$\Dalpha$, which plays an important role in constraining the Standard
Model (SM) in the context of electroweak precision measurements, so that
its precise theoretical understanding is an important task.

The parameter $\Dalpha$ receives leptonic and hadronic contributions.
The latter escape a perturbative treatment and are determined using
cross-section measurements of electron positron annihilation into
hadrons, see
e.g. Refs.~\cite{Kuhn:1998ze,Martin:2000by,deTroconiz:2004tr,Burkhardt:2005se,Davier:2010nc,Jegerlehner:2011mw,Hagiwara:2011af,Bodenstein:2012pq}.
The hadronic contributions $\Dalpha_{\mbox{\scriptsize{had}}}$
constitute currently the dominant uncertainty to $\Dalpha$. The leptonic
contributions $\Dalpha_{\mbox{\scriptsize{lep}}}$ at one- and two-loop
order in QED are known from the results for the vacuum polarization
function of Refs.~\cite{Kallen:1955fb,Schwinger:1989ka}. The three-loop
order has been computed in Ref.~\cite{Steinhauser:1998rq} in terms of
expansions in the small mass ratio $M_{\ell}^2/M_Z^2$, where $M_{\ell}$
stands for the masses of the SM charged leptons and $M_Z$ is the
$Z$-boson mass. The dominant contribution to
$\Dalpha_{\mbox{\scriptsize{lep}}}$ arises form large logarithms in this
small mass ratio, so that it is often sufficient to perform the
expansion up to the constant term, including it, and to neglect power
suppressed terms of higher order in the mass ratio $M_{\ell}^2/M_Z^2$.
Within this work we will extend the calculation of
$\Dalpha_{\mbox{\scriptsize{lep}}}$ to four-loop order in QED, also in
form of small lepton mass expansions up to the constant term. For the
lower orders in perturbation theory we include the power suppressed
terms.  Next to the effective electromagnetic coupling we will also
present an numerical evaluation of $\Dalpha_{\mbox{\scriptsize{lep}}}$,
which allows to study the size of the different contributions.  As a
by-product of our calculation we will also provide the result for the
relation between the $\MSbar$ and on-shell mass from the contribution
due to heavy fermion loops at three-loop order in QED.

In the next Section~\ref{sec:Generalities} we start with some
generalities and the definition of our notation. In
Section~\ref{sec:Calculation} we give an outline of our calculation.
The new results for the vacuum polarization function in the on-shell
scheme at four-loop order are presented in Section~\ref{sec:Results}
including a numerical evaluation of $\Dalpha_{\mbox{\scriptsize{lep}}}$.
Finally we close with our summary and conclusions in
Section~\ref{sec:DiscussConclude}.  In the Appendix we provide known
results as supplementary information.
\section{Generalities and notation\label{sec:Generalities}}
We consider the photon vacuum polarization in QED with all one particle
irreducible insertions into the photon propagator. The vacuum
polarization tensor is given by
\begin{equation}
\Pi^{\mu\nu}(q,M)=i\!\int\!d x\,e^{i q x} \langle0|Tj^{\mu}(x)j^{\nu}(0)|0\rangle,
\end{equation}
where $j^{\mu}(x)$ is the electromagnetic current.  It can be decomposed
with respect to its Lorentz structure into two terms
\begin{eqnarray}
\Pi^{\mu\nu}\!\left(q,M\right)&=&
  \left(-q^2\*g^{\mu\nu}+q^{\mu}\*q^{\nu}\right)\*\Pi\left(q^2,M\right) 
 + q^{\mu}\*q^{\nu}\*\Pi_{L}\left(q^2,M\right),
\end{eqnarray}
where $\Pi\left(q^2,M\right)$ and $\Pi_{L}\left(q^2,M\right)$ are the
transversal and longitudinal parts of the vacuum polarization.  The
longitudinal part vanishes due to the Ward-Takahashi identities.  In the
following the symbols $M_e$, $M_\mu$, $M_\tau$ denote the on-shell
masses of the electron(e), muon($\mu$) and tauon($\tau$) with the mass
hierarchy $M_e<\!\!<M_\mu<\!\!<M_\tau$.  The polarization function
$\Pi\left(q^2,M\right)$ depends on the lepton masses
$M=\{M_e,M_\mu,M_\tau\}$ as well as on the external Minkowskian momentum
$q$, where we consider here $q^2>0$.  It is related to the effective
electromagnetic coupling by
\begin{equation}
\alpha_{\mbox{\scriptsize{eff}}}(q^2)={\alpha\over 1 + \mbox{Re}\Pi_{\os}(q^2)}\,,
\end{equation}
which describes the running of the electromagnetic coupling to higher
energy scales. The running from $q^2=0$ to the scale of the electroweak
$Z$-boson mass, $q^2=M_Z^2$, is mediated by the parameter $\Dalpha$,
where we consider within this work the purely leptonic contributions,
\begin{equation}
\label{eq:DefDeltaAlpha}
\Dalpha_{\mbox{\scriptsize{lep}}}\equiv\Dalpha_{\mbox{\scriptsize{lep}}}(M_Z^2)=-\mbox{Re}\Pi_{\os}(q^2=M_Z^2)\,,
\end{equation}
from the Standard Model leptons which we defined above. The symbol Re
denotes the real part of a complex quantity and the label $\os$
specifies that the vacuum polarization function has been renormalized in
the on-shell scheme. In the on-shell scheme holds $\Pi_{\os}(q^2=0)=0$.\\
In order to determine $\Dalpha_{\mbox{\scriptsize{lep}}}$ we need to
compute the vacuum polarization function in the on-shell scheme. We
define its perturbative expansion as
\begin{equation}
\label{eq:Piexp}
\Pi_{\os}(q^2,M)={\aemOSvpi}\*\sum_{k=0}
 \left(\aemOSpi\right)^{k}\*\Pi_{\os}^{(k)}(q^2,M)\,,
\end{equation}
where each term on the r.h.s. of Eq.~(\ref{eq:Piexp}) stands for the
one-loop($k=0$), two-loop($k=1$), three-loop($k=2$) and four-loop($k=3$)
contribution.  The vacuum polarization function in the on-shell scheme
at one- and two-loop order is known since
long~\cite{Kallen:1955fb,Schwinger:1989ka}.  The one-loop result reads
\begin{eqnarray}
\label{eq:Pi0}
\Pi_{\os}^{(0)}(q^2,M)&=&\!\!\!\!\!
\sum_{i={e,\mu,\tau}}\!\!
   {4\over3}\*\left[
   {5\over3}
 + x_i
 - (2+x_i)\*{\beta_i\over2}\*
   \log{\left({\beta_i+1\over\beta_i-1}\right)}
             \right] 
\\\label{eq:Pi0exp}
&\stackrel{q^2>\!\!>(2\*M_i)^2}{\approx}&\!\!\!\!\!
\sum_{i={e,\mu,\tau}}\!\!\left[
 - {4\over3}\*\LL{i}
 + {20\over 9} 
 + 8\*{M_i^2\over q^2} 
 + 4\*\left({M_i^2\over q^2}\right)^2\!\*
   \left(2\*\LL{i}+1\right)
+\dots\!\right]\!,
\end{eqnarray}
with $\beta_i^2=1 - x_i$, $x_i=(2\*M_i)^2/q^2$ and $\LL{i}=\log{(-{q^2 /
    M_i^2})}$. The ellipsis in Eq.~(\ref{eq:Pi0exp}) represent the
contributions from higher orders in the expansion in $M_i^2/q^2$ for an
external momentum $q^2$ which is much larger than the threshold for
lepton pair production. The index $i$ stands for the flavor of the
lepton which is connected to the external photons. This notation will be
kept also in the subsequent equations.  The analytic continuation of the
logarithm is determined by the $i\ep$-prescription.\\
The two-loop result is given by
\begin{eqnarray}
\label{eq:Pi1}
\Pi_{\os}^{(1)}(q^2,M)\!&=&\!\!\!\!\sum_{i={e,\mu,\tau}}\!\left[
- \LL{i} 
+ {5\over 6} - 4\*\z3 
- 12\*{M_i^2\over q^2}\*\LL{i}
\right.\nonumber\\&&\left.\qquad\quad\!
- \left({M_i^2\over q^2}\right)^2\*\left( 12\*\LL{i}^2 + 10\*\LL{i} - {2\over3} - 16\*\z3 \right)
+ \mathcal{O}\left(M_i^6\over q^4\right)\right]\!.
\end{eqnarray}

\begin{figure}[!ht]
\begin{center}
\begin{minipage}{3cm}
\begin{center}
\includegraphics[bb=97 433 379 675,width=3cm]{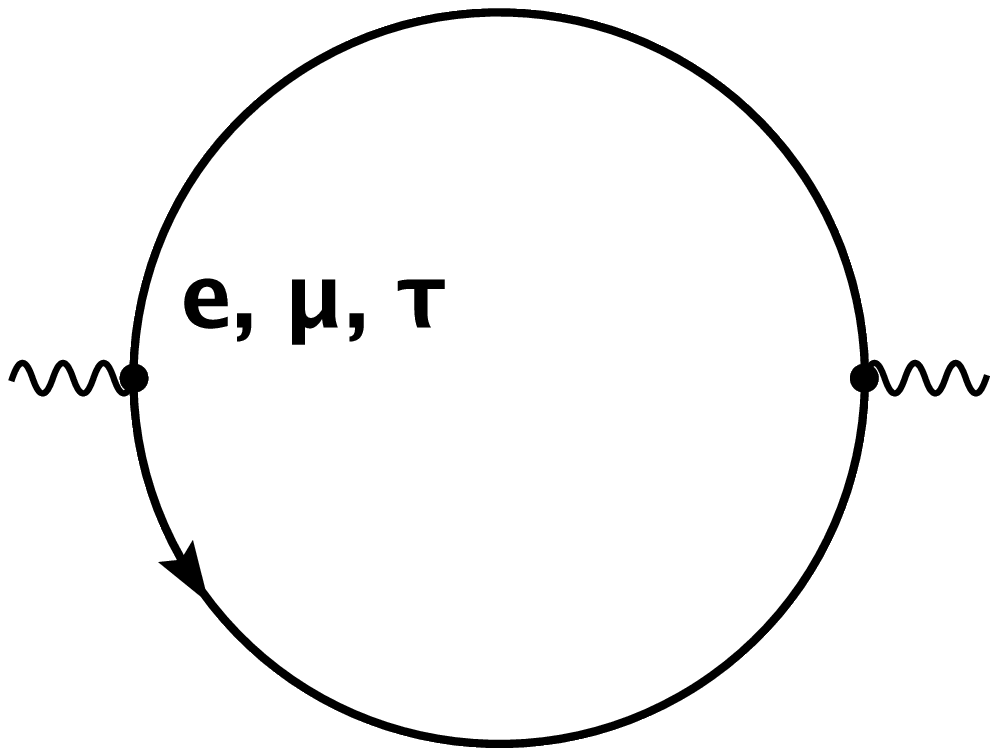}\\
(a)
\end{center}
\end{minipage}
\hspace{0.5cm}
\begin{minipage}{3cm}
\begin{center}
\includegraphics[bb=97 433 379 675,width=3cm]{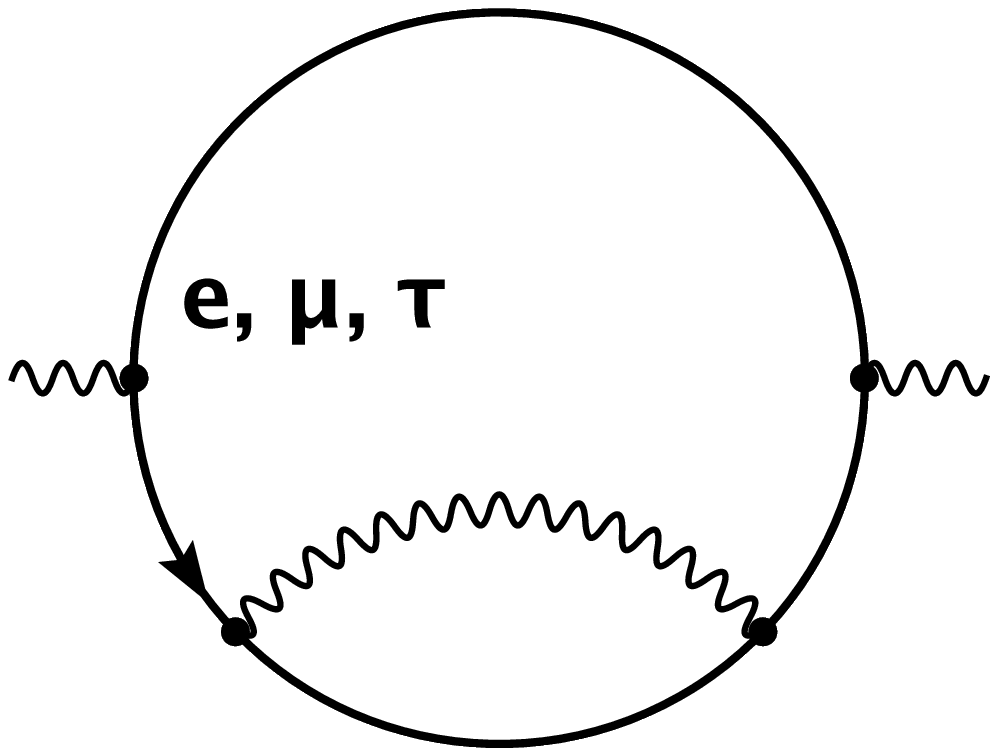}\\
($\mbox{b}_1$)
\end{center}
\end{minipage}
\hspace{0.2cm}
\begin{minipage}{3cm}
\begin{center}
\includegraphics[bb=97 433 379 675,width=3cm]{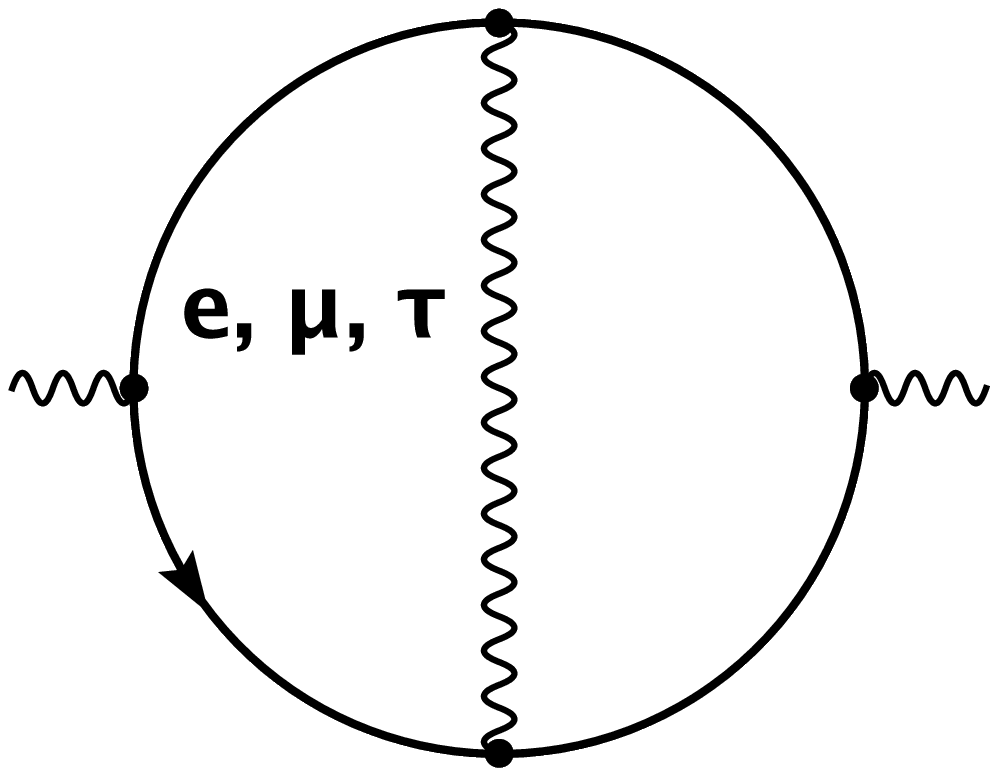}\\
($\mbox{b}_2$)
\end{center}
\end{minipage}
\hspace{0.2cm}
\begin{minipage}{3cm}
\begin{center}
\includegraphics[bb=97 433 379 675,width=3cm]{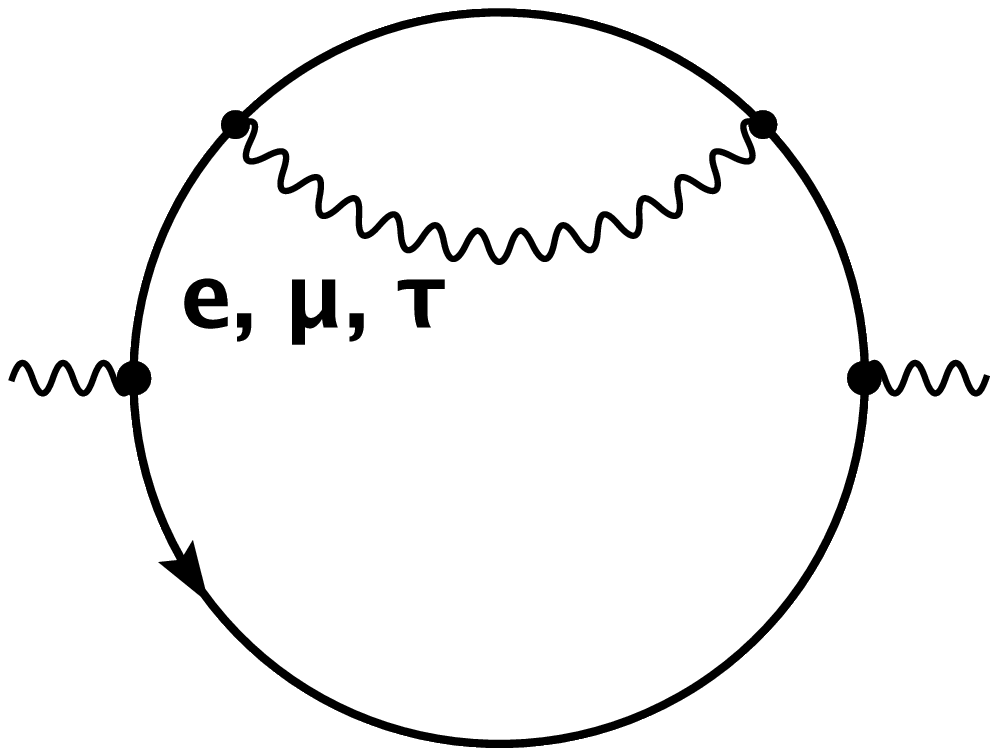}\\
($\mbox{b}_3$)
\end{center}
\end{minipage}
\end{center}
\vspace*{-0.4cm}
\caption{Diagrams contributing to the one-loop and two-loop vacuum
  polarization function. The solid lines denote charged leptons ($e$, $\mu$,
  $\tau$); the wavy lines denote photons.\label{fig:12loop}}
\end{figure}

\begin{figure}[!ht]
\begin{center}
\begin{minipage}{3cm}
\begin{center}
\includegraphics[bb=97 433 379 675,width=3cm]{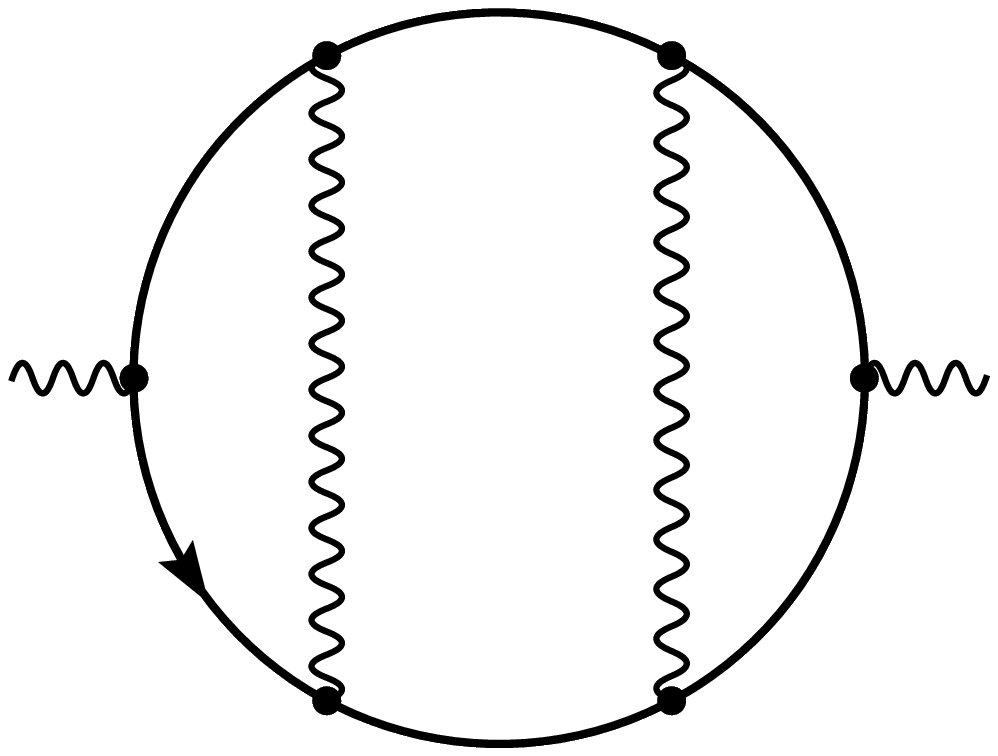}\\[-0.1cm]
\hspace{2ex}$\Pi_{A}^{(2)}$
\end{center}
\end{minipage}
\hspace{0.3cm}
\begin{minipage}{3cm}
\begin{center}
\includegraphics[bb=97 433 379 675,width=3cm]{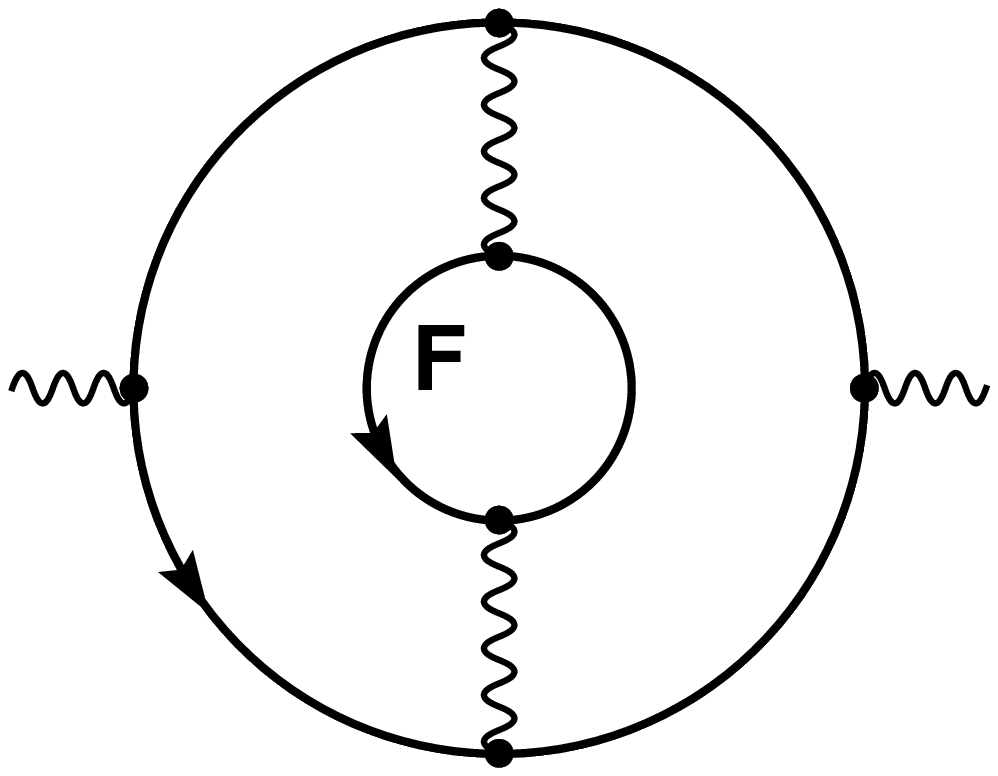}\\[-0.1cm]
\hspace{2ex}$\Pi_{F}^{(2)}$
\end{center}
\end{minipage}
\hspace{0.3cm}
\begin{minipage}{3cm}
\begin{center}
\includegraphics[bb=97 433 379 675,width=3cm]{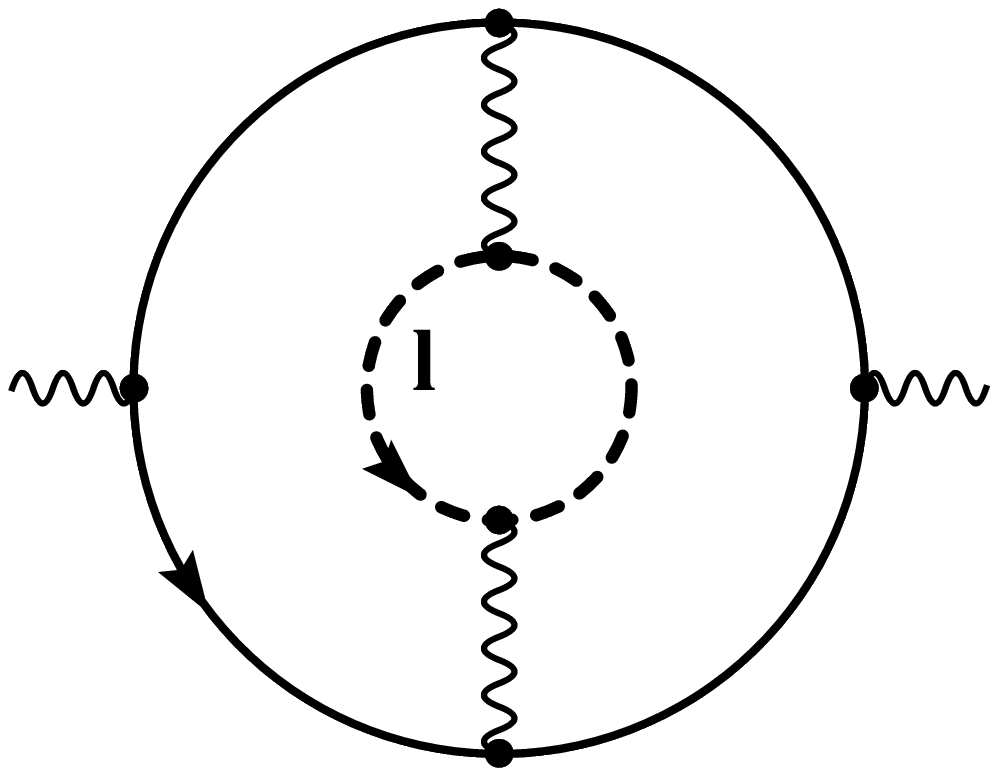}\\[-0.1cm]
\hspace{2ex}$\Pi_{l}^{(2)}$
\end{center}
\end{minipage}
\hspace{0.3cm}
\begin{minipage}{3cm}
\begin{center}
\includegraphics[bb=97 433 379 675,width=3cm]{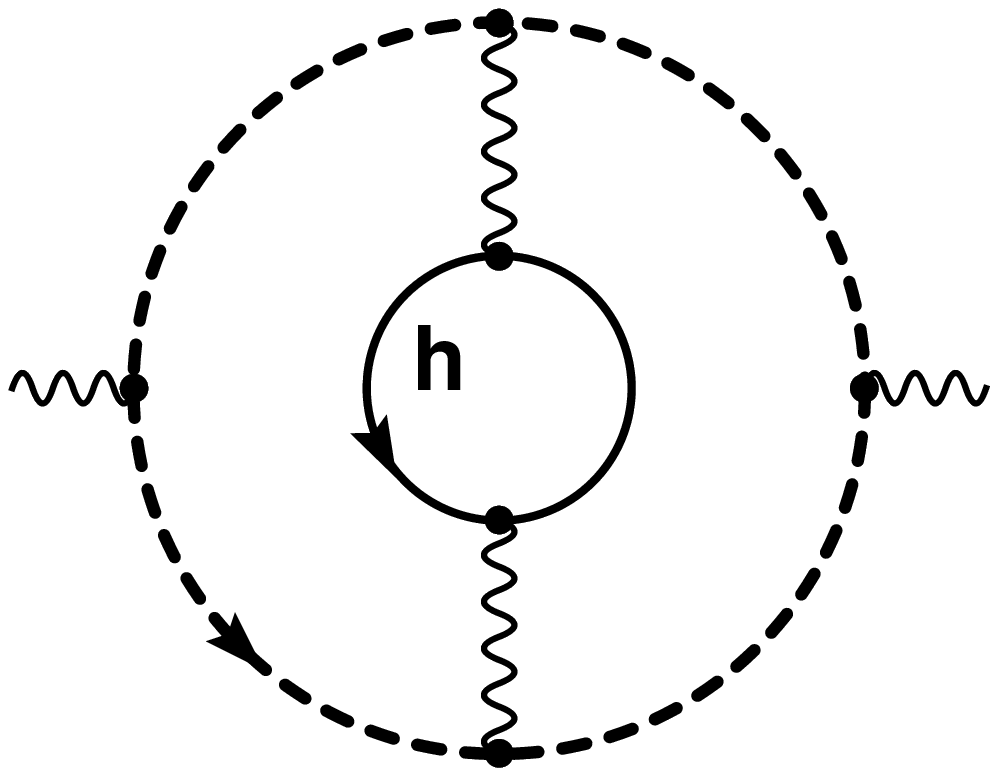}\\[-0.1cm]
\hspace{2ex}$\Pi_{h}^{(2)}$
\end{center}
\end{minipage}
\end{center}
\vspace*{-0.4cm}
\caption{Examples for the different diagram classes at three-loop
  order. For each class one typical representative is shown. Loops of
  particles which are represented by a dashed line denote particles
  which have a lower mass than the ones which are represented by a solid
  line. \label{fig:3loop}}
\end{figure}

The diagrams which contribute to the one- and two-loop vacuum
polarization function are shown in Fig.~\ref{fig:12loop}.\\
\\
The corresponding result in QCD, where quarks instead of leptons are
running in the fermion loop, can easily be deduce up to two-loop order
from Eqs.~(\ref{eq:Pi0}) and (\ref{eq:Pi1}) by multiplying them with the
appropriate color factors.\\
The three-loop contribution has been determined in terms of expansions
in the low ($q^2\to0$) and the high energy limit in
Refs.~\cite{Gorishnii:1990vf,Surguladze:1990tg,Broadhurst:1992za,Chetyrkin:1994ex,Chetyrkin:1995ii,Baikov:1995ui,Chetyrkin:1996cf,Chetyrkin:1996ez,Chetyrkin:1997qi}. The
result in the on-shell scheme has been computed in
Ref.~\cite{Steinhauser:1998rq}, where it has been decomposed into four
contributions
\begin{eqnarray}
\label{eq:3loop}
\Pi_{\os}^{(2)}(q^2,M)&=&\sum_{i={e,\mu,\tau}}\!\left(
   \Pi_{A}^{(2)}(q^2,M_i)
 + \Pi_{F}^{(2)}(q^2,M_i)
                                     \right)\nonumber\\
&+&\sum_{\substack{i={\mu,\tau}\\j={e,\mu}\\ j\ne i}}\Pi_{l}^{(2)}(q^2,M_i,M_j)
 + \sum_{\substack{i={e,\mu}\\j={\mu,\tau}\\ j\ne i}}\Pi_{h}^{(2)}(q^2,M_i,M_j).
\end{eqnarray}
The first term $\Pi_{A}^{(2)}(q^2,M_i)$ is the quenched contribution
which originates from diagrams without any insertion of an internal,
closed fermion loop. In contrast, the terms $\Pi_{F}^{(2)}(q^2,M_i)$,
$\Pi_{l}^{(2)}(q^2,M_i,M_j)$ and $\Pi_{h}^{(2)}(q^2,M_i,M_j)$ arise from
diagrams with internal lepton loop insertions. The contribution with the
subscript $F$ arises from diagrams with an internal lepton loop of the
same flavor as the outer loop which is connected to the external
photons; the contribution with the subscript $h$ comes from diagrams
with a {\underline{h}}eavier lepton inserted in the inner loop than in
the outer one, and the subscript $l$ denotes the contribution where the
mass of the lepton in the inner loop is {\underline{l}}ighter than the
external one.  An example-diagram for each of these four different kind
of contributions is shown in Fig.~\ref{fig:3loop}.\\
\\
The four different contributions on the r.h.s. of Eq.~(\ref{eq:3loop})
are renormalized in the on-shell scheme and we omit a label indicating
it for brevity.  The index $i$ labels again the external leptons which
connect to the external photons in a diagram, whereas the index $j$
enumerates the internal leptons. The three-loop on-shell results for the
different terms on the r.h.s. of Eq.~(\ref{eq:3loop}) were determined in
terms of expansions in the limit $q^2\!>\!\!>\!(2\*M)^2$ in
Ref.~\cite{Steinhauser:1998rq} and are shown for completeness in
Appendix~\ref{app:Pi3loop}, where we have in addition also added mass
corrections of higher orders in the small ratio $M^2/q^2$.  For
$\Pi_{A}^{(2)}(q^2,M_i)$ and $\Pi_{F}^{(2)}(q^2,M_i)$ they can be
obtained from the $\MSbar$ results in
Refs.~\cite{Chetyrkin:1994ex,Chetyrkin:1995ii,Chetyrkin:1996cf,Chetyrkin:1997qi}
after converting them from the $\MSbar$ to the on-shell scheme.

The four-loop result can be decomposed, similarly to
Eq.~(\ref{eq:3loop}), into several gauge invariant contributions
according to the properties and number of the inserted lepton loops
\begin{eqnarray}
\label{eq:pi4loop}
\Pi_{\os}^{(3)}(q^2,M)&=&
  \sum_{i=e,\mu,\tau}\left[
   \Pi_{A }^{(3)}(q^2,M_i)
 + \Pi_{FF}^{(3)}(q^2,M_i)
 + \Pi_{F }^{(3)}(q^2,M_i)
                  \right]
\nonumber\\
&+&\sum_{\substack{i={e,\mu}\\j={\mu,\tau}\\ j\ne i}}\!\left[
   \Pi_{h}^{(3)}(q^2,M_j)
 + \Pi_{Fh}^{(3)}(q^2,M_i,M_j)
 + \Pi_{hh}^{(3)}(q^2,M_j)
                                            \right]
\nonumber\\
&+&\sum_{\substack{i={\mu,\tau}\\j={e,\mu}\\ j\ne i}}\!\left[
   \Pi_{l}^{(3)}(q^2,M_i,M_j)
 + \Pi_{Fl}^{(3)}(q^2,M_i,M_j)
 + \Pi_{ll}^{(3)}(q^2,M_i,M_j)
                                            \right]
\nonumber\\
&+&\Pi_{e\mu}^{(3)}(q^2,M_\tau,M_e,M_\mu) 
 + \sum_{j={e,\mu}}\!\Pi_{j\tau}^{(3)}(q^2,M_j,M_\tau) 
\nonumber\\
&+&\sum_{i=e,\mu,\tau}\!\Pi_{s,d}^{(3)}(q^2,M_i)
 + \sum_{\substack{k,l=e,\mu,\tau\\k\ne l}}\!\Pi_{s,nd}^{(3)}\left(q^2,\mbox{max}\{M_k,M_l\}\right)\,.
\end{eqnarray}
The property which characterizes each individual term on the r.h.s. of
Eq.~(\ref{eq:pi4loop}) is defined in Fig.~\ref{fig:dia} by the shown
example-diagrams. Singlet diagrams, in which the external photons are
not connected to the same fermion loop did not yet contribute at
three-loop order due to Furry's theorem\cite{Furry:1937zz} and survive
for the first time at four-loop order. They are given by the terms
$\Pi_{s,d}^{(3)}(q^2,M_i)$ and
$\Pi_{s,nd}^{(3)}\left(q^2,\mbox{max}\{M_k,M_l\}\right)$, where the
subscript {\footnotesize{$d$}} stands for `{\underline{d}}iagonal' which
denotes diagrams with the same lepton flavor in the two fermion loops,
whereas the subscript {\footnotesize{$nd$}} stands for
`{\underline{n}}on-{\underline{d}}iagonal' and denotes diagrams with
different lepton flavors in the two fermion loops.  The diagrams with
three different lepton flavors which correspond to the three terms
$\Pi_j^{(3)}$ with $j=\{e\mu, e\tau,\mu\tau\}$ of Eq.~(\ref{eq:pi4loop}),
also arise for the first time at four-loop order.  As in the previous
Eq.~(\ref{eq:3loop}) the subscripts
{\footnotesize{$\{FF,F,h,Fh,hh,l,Fl,ll\}$}} of the non-singlet
contributions in Eq.~(\ref{eq:pi4loop}) refer always to the internal
leptons. Apart from that we show only the function arguments of the
lepton masses on which these expanded results will depend to the
considered expansion depth. In general, without expansion, the functions
may depend on more mass arguments than those which are given here.  The
on-shell results for $\Pi_{A}^{(3)}(q^2,M_i)$,
$\Pi_{FF}^{(3)}(q^2,M_i)$, $\Pi_{F}^{(3)}(q^2,M_i)$ and
$\Pi_{s,d}^{(3)}(q^2,M_i)$ have already been determined in
Refs.~\cite{Baikov:2008si,Baikov:2012rr} in the high energy expansion.
They are given for completeness in Appendix~\ref{app:Pi4loopOld}.  The
calculation of the terms in Eq.~(\ref{eq:pi4loop}) which are still
unknown in order to describe the complete lepton mass hierarchy will be
discussed in the next section.
\section{Calculation\label{sec:Calculation}}
In order to compute the leading four-loop QED corrections to
$\Delta\alpha_{\mbox{\scriptsize{lep}}}$ we will determine the on-shell
vacuum polarization function contribution $\Pi_{\os}^{(3)}(q^2=M_Z^2,M)$
of Eq.~(\ref{eq:pi4loop}) in the small lepton mass expansion up to power
corrections of the order $M^2/M_Z^2$.  For this purpose we need to know
the $\MSbar$ vacuum polarization function on the one hand for massless
leptons and on the other hand we need to know it in the low energy limit
at $q^2=0$ for the renormalization procedure in the on-shell scheme as
we will see in the following.

The vacuum polarization function in the massless limit,
$\overline{\Pi}^{(k)}(q^2, \overline{m}=0)$, is known analytically in
the $\MSbar$ scheme up to four-loop
order($k=3$)~\cite{Gorishnii:1990vf,Surguladze:1990tg,Chetyrkin:1996ez,Baikov:2008si,Baikov:2009uw,Baikov:2012rr,
  Baikov:2012zm,Baikov:2012zn}.
In this limit it does not depend on the lepton masses but only on the
external momentum. The perturbative expansion of the vacuum polarization
function in the $\MSbar$ scheme is defined here in complete analogy to
the on-shell case shown in Eq.~(\ref{eq:Piexp}), only that all
quantities are renormalized in the $\MSbar$ scheme.

\begin{figure}[!ht]
\begin{center}
\begin{minipage}{3cm}
\begin{center}
\includegraphics[bb=97 434 379 674,width=3cm]{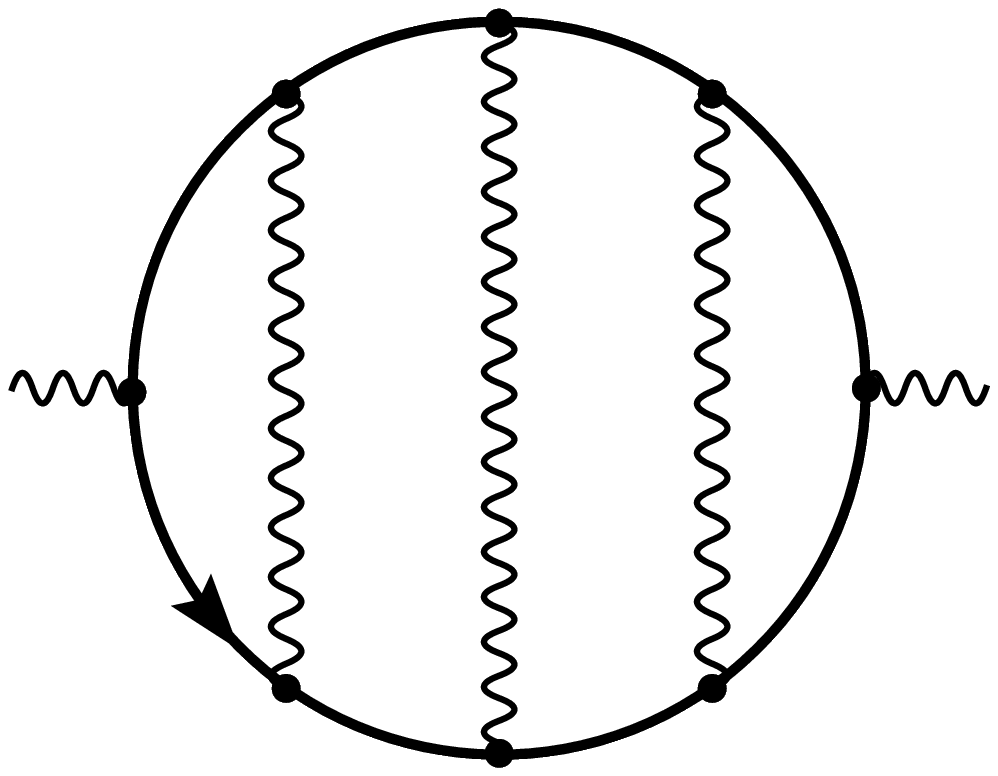}\\[-0.1cm]
\hspace{2ex}$\Pi_{A}^{(3)}$
\end{center}
\end{minipage}
\hspace{1cm}
\begin{minipage}{3cm}
\begin{center}
\includegraphics[bb=97 437 379 671,width=3cm]{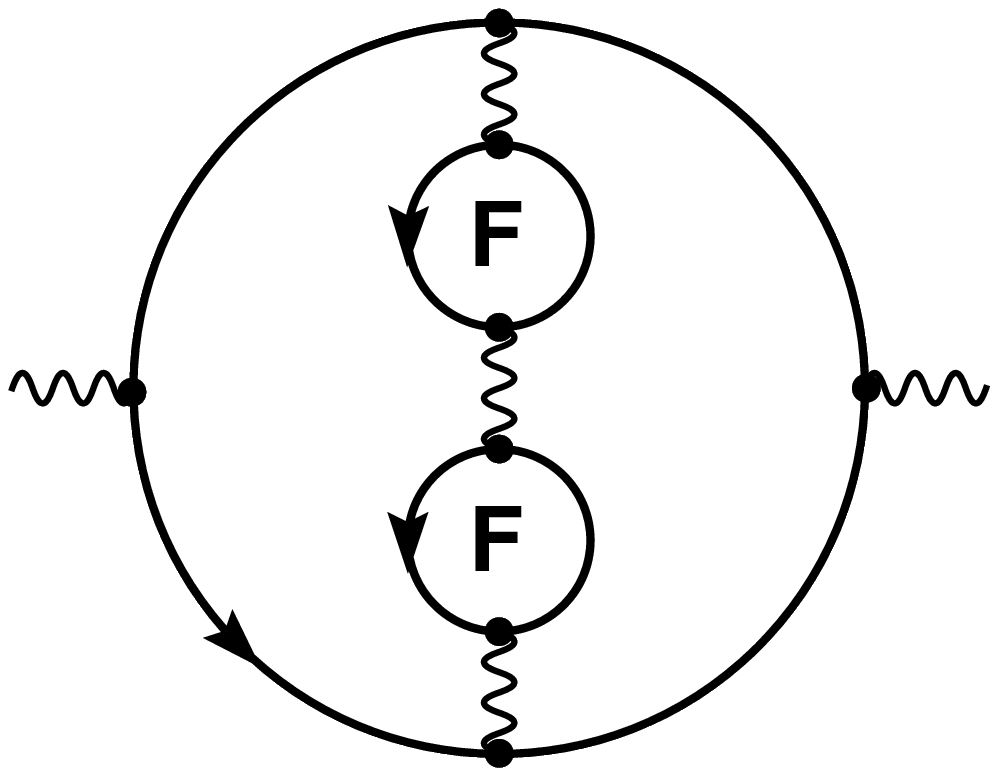}\\[-0.1cm]
\hspace{2ex}$\Pi_{FF}^{(3)}$
\end{center}
\end{minipage}
\hspace{1cm}
\begin{minipage}{3cm}
\begin{center}
\includegraphics[bb=97 437 379 671,width=3cm]{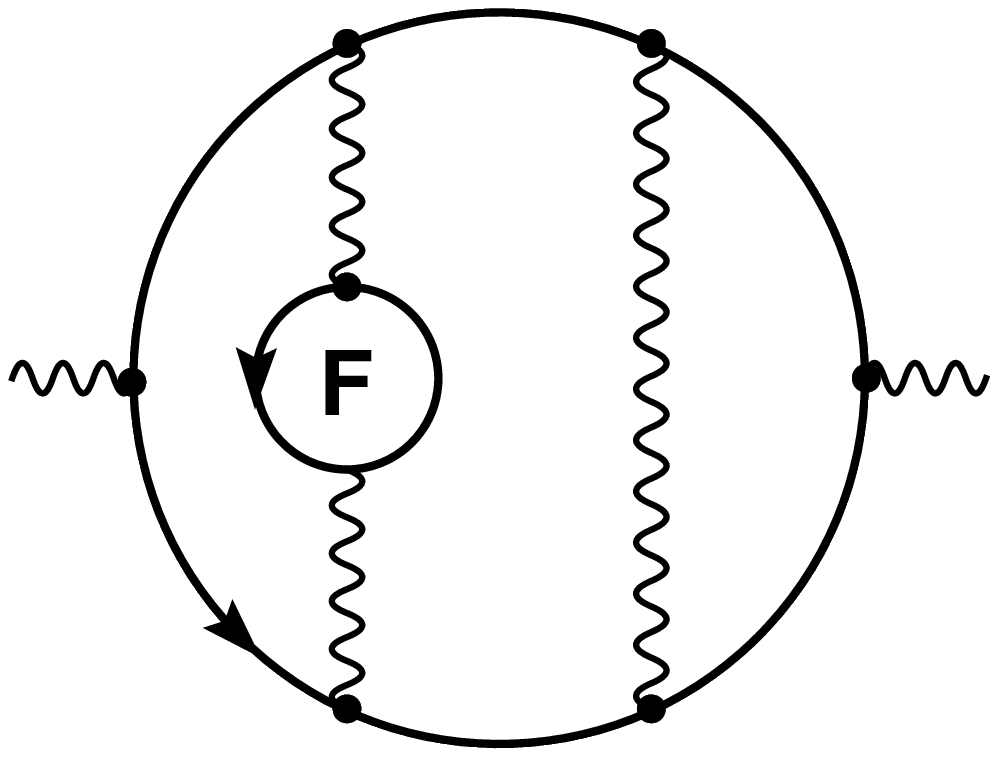}\\[-0.1cm]
\hspace{2ex}$\Pi_{F}^{(3)}$
\end{center}
\end{minipage}\\[0.2cm]
\begin{minipage}{3cm}
\begin{center}
\includegraphics[bb=97 447 379 661,width=3cm]{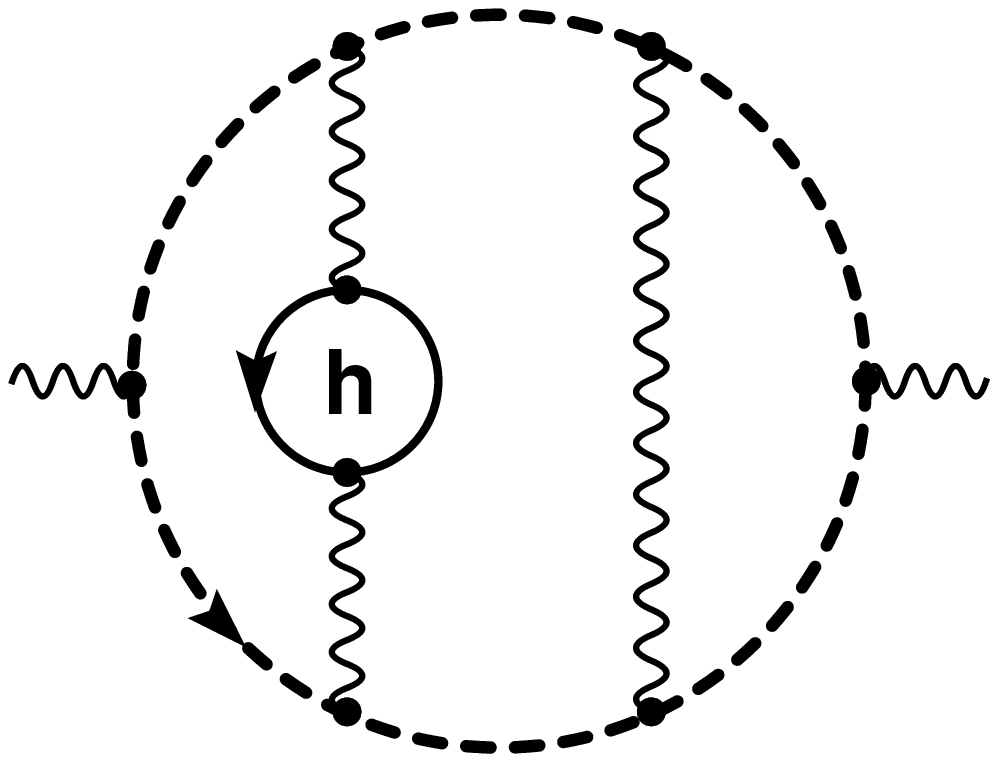}\\[-0.0cm]
\hspace{2ex}$\Pi_{h}^{(3)}$
\end{center}
\end{minipage}
\hspace{1cm}
\begin{minipage}{3cm}
\begin{center}
\includegraphics[bb=97 443 379 665,width=3cm]{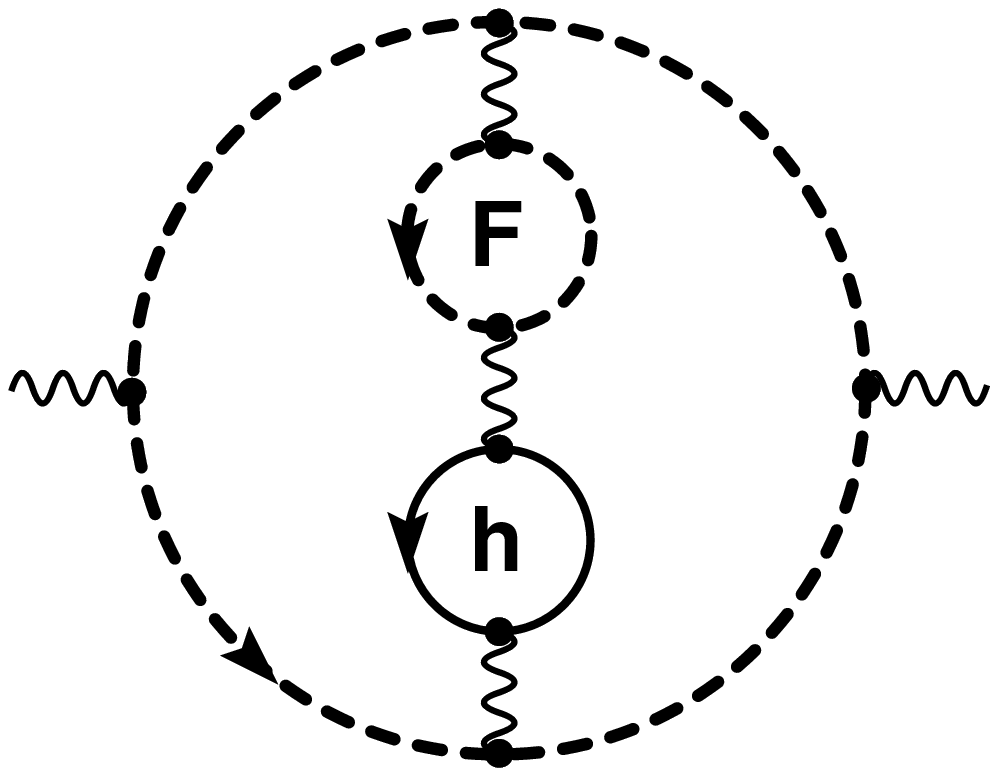}\\[-0.0cm]
\hspace{2ex}$\Pi_{Fh}^{(3)}$
\end{center}
\end{minipage}
\hspace{1cm}
\begin{minipage}{3cm}
\begin{center}
\includegraphics[bb=97 443 379 665,width=3cm]{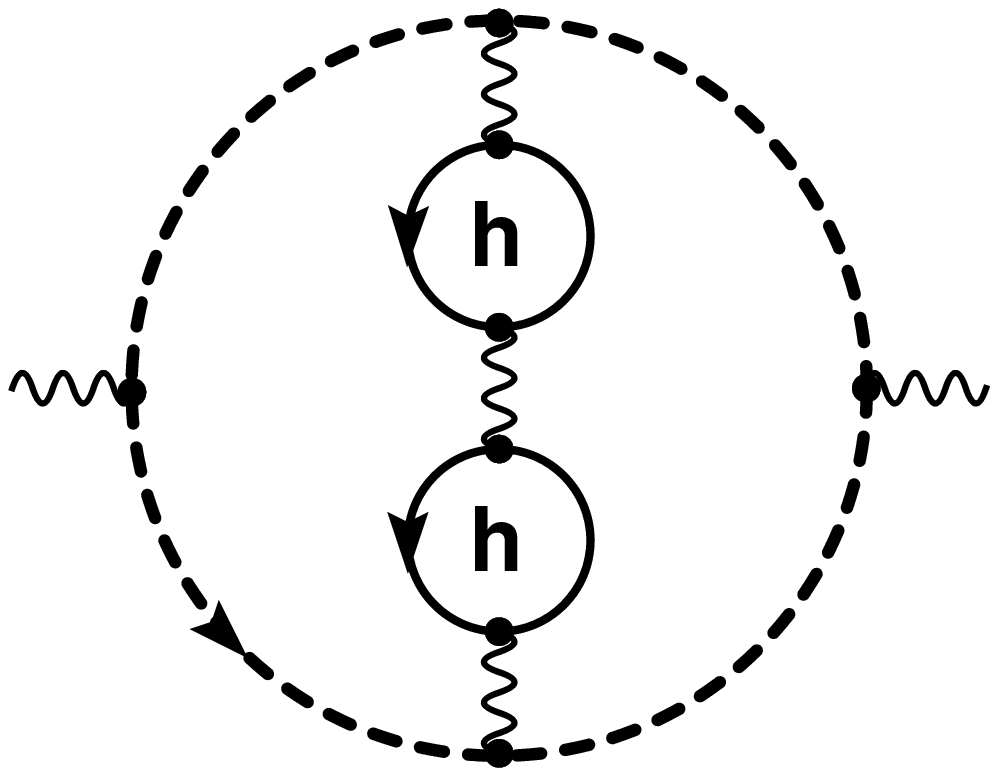}\\[-0.0cm]
\hspace{2ex}$\Pi_{hh}^{(3)}$
\end{center}
\end{minipage}\\[0.2cm]
\begin{minipage}{3cm}
\begin{center}
\includegraphics[bb=97 437 379 671,width=3cm]{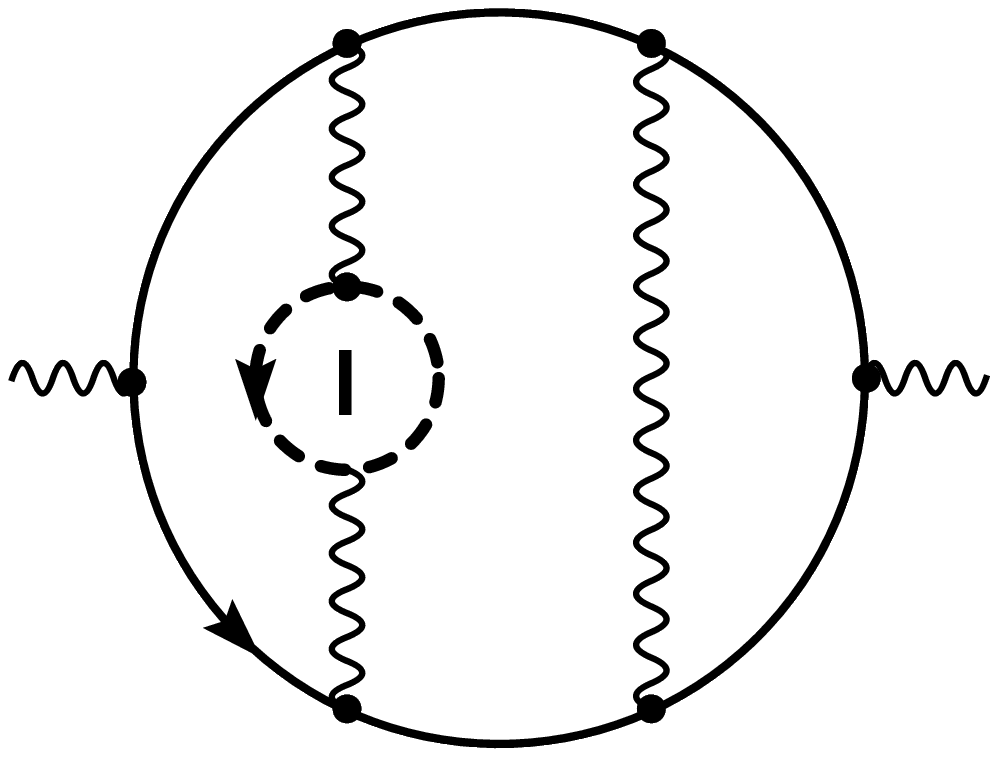}\\[-0.1cm]
\hspace{2ex}$\Pi_{l}^{(3)}$
\end{center}
\end{minipage}
\hspace{1cm}
\begin{minipage}{3cm}
\begin{center}
\includegraphics[bb=97 437 379 671,width=3cm]{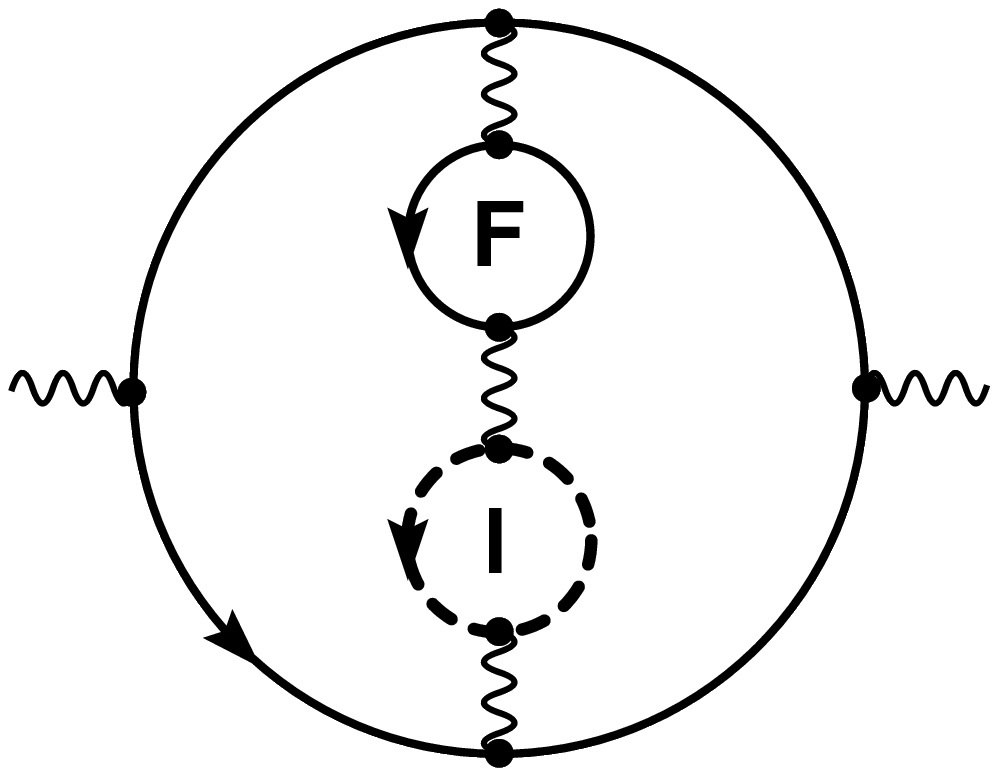}\\[-0.1cm]
\hspace{2ex}$\Pi_{Fl}^{(3)}$
\end{center}
\end{minipage}
\hspace{1cm}
\begin{minipage}{3cm}
\begin{center}
\includegraphics[bb=97 437 379 671,width=3cm]{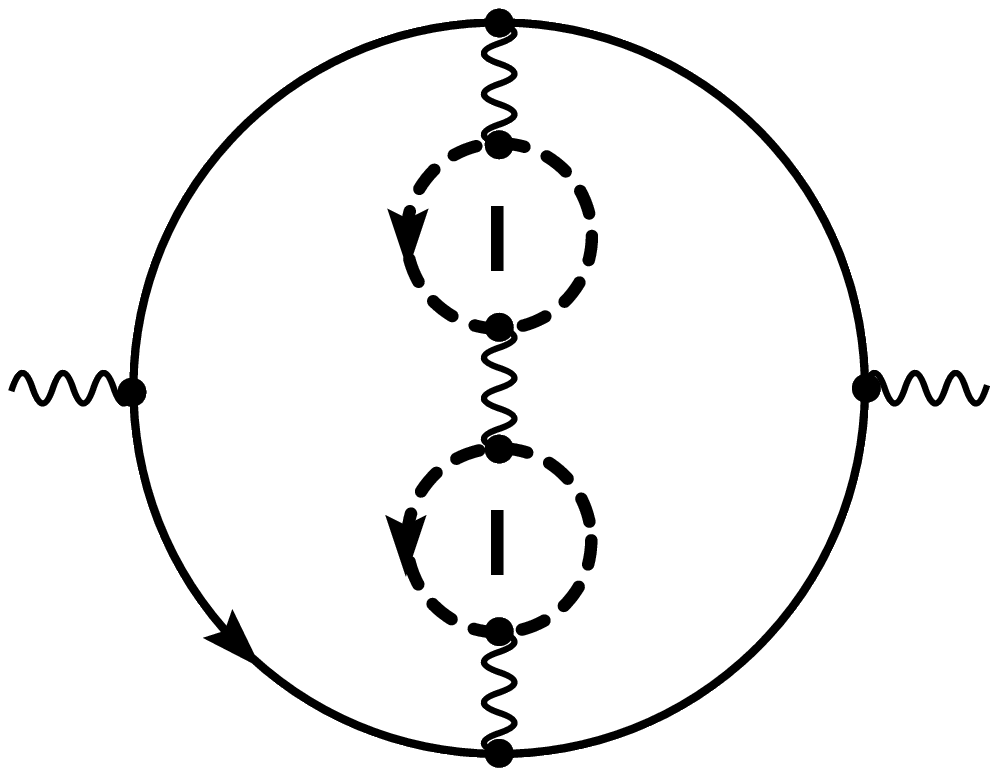}\\[-0.1cm]
\hspace{2ex}$\Pi_{ll}^{(3)}$
\end{center}
\end{minipage}\\[0.2cm]
\begin{minipage}{3cm}
\begin{center}
\includegraphics[bb=97 434 379 674,width=3cm]{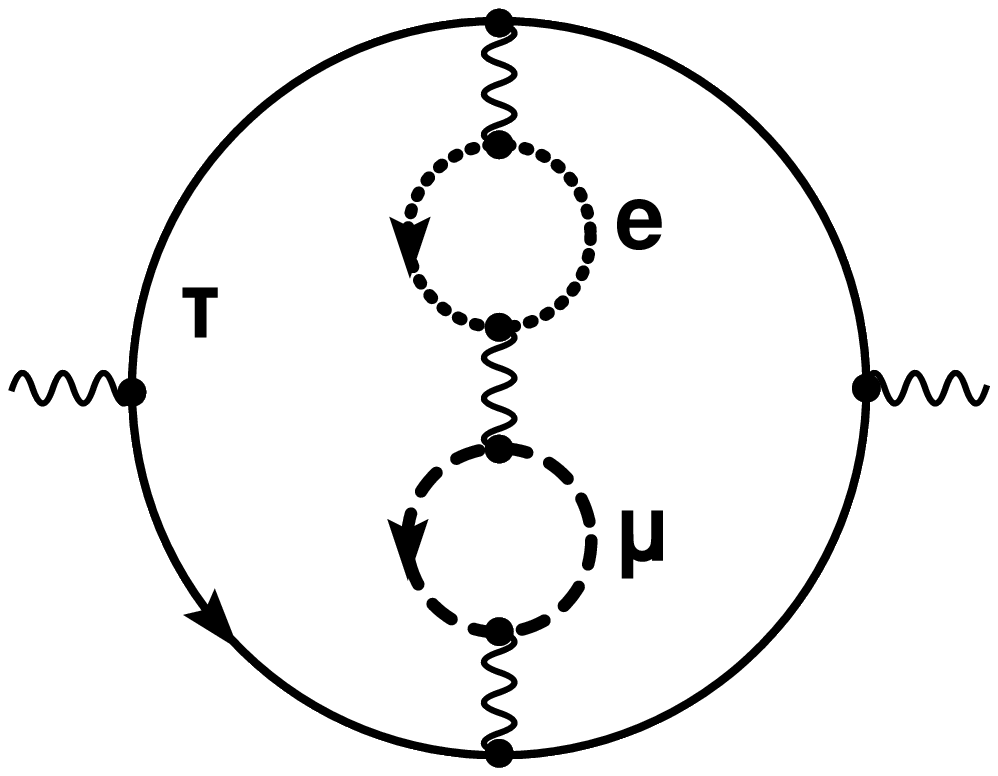}\\[-0.1cm]
\hspace{2ex}$\Pi_{e\mu}^{(3)}$
\end{center}
\end{minipage}
\hspace{1cm}
\begin{minipage}{3cm}
\begin{center}
\includegraphics[bb=97 443 379 665,width=3cm]{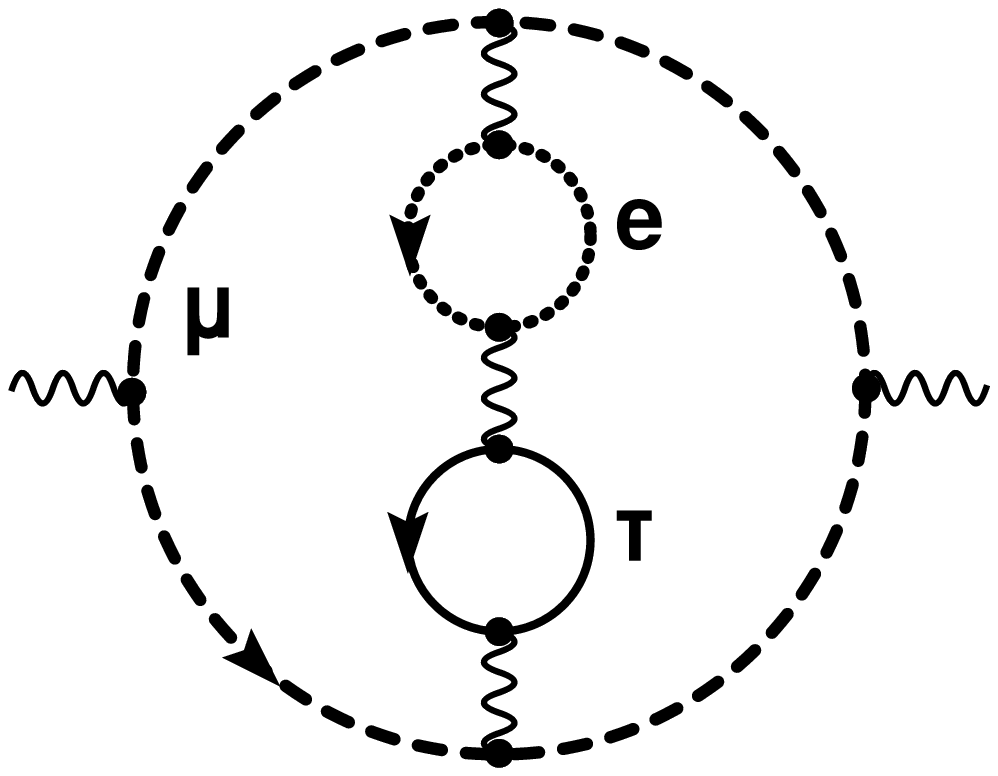}\\[-0.1cm]
\hspace{2ex}$\Pi_{e\tau}^{(3)}$
\end{center}
\end{minipage}
\hspace{1cm}
\begin{minipage}{3cm}
\begin{center}
\includegraphics[bb=97 443 379 665,width=3cm]{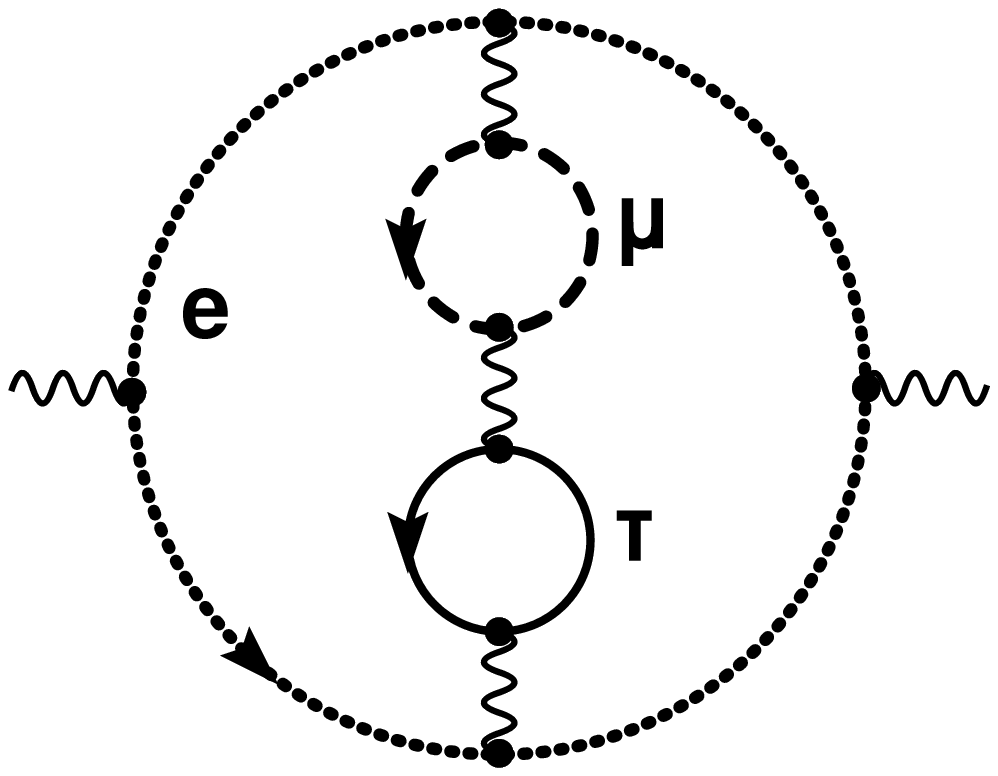}\\[-0.1cm]
\hspace{2ex}$\Pi_{\mu\tau}^{(3)}$
\end{center}
\end{minipage}\\[-0.1cm]
\begin{minipage}{3cm}
\begin{center}
\raisebox{0.1cm}{
\includegraphics[bb=97 441 379 666,width=3cm]{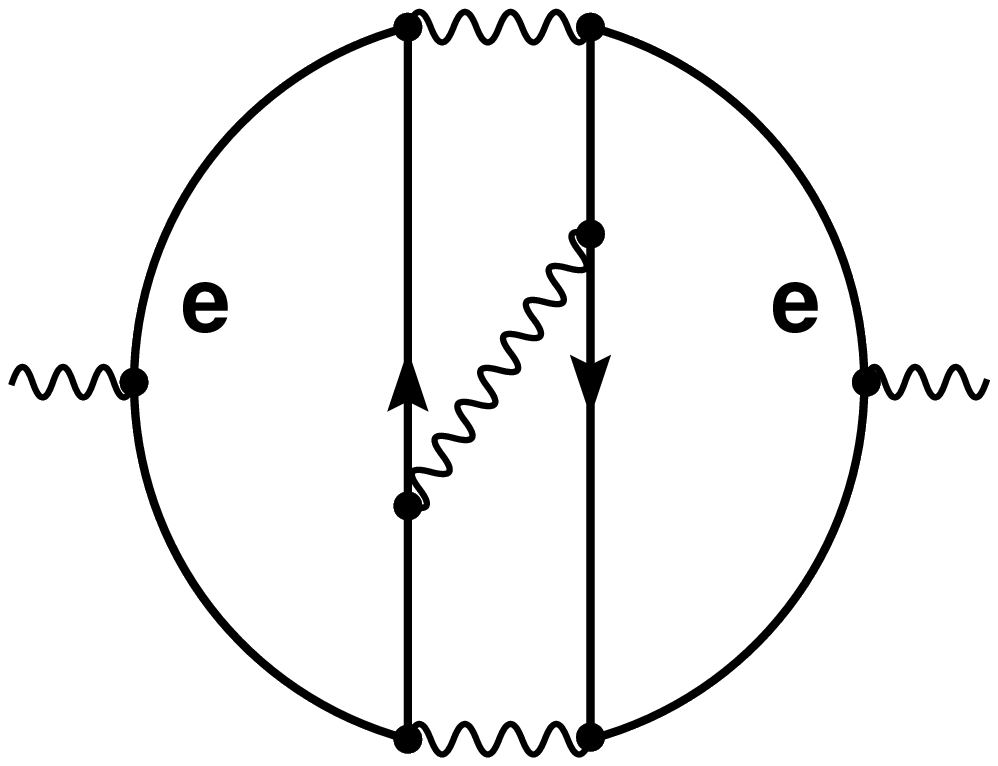}
           }\\[-0.1cm]
\hspace{2ex}$\Pi_{s,d}^{(3)}$
\end{center}
\end{minipage}
\hspace{1cm}
\begin{minipage}{3cm}
\begin{center}
\raisebox{0.1cm}{
\includegraphics[bb=97 441 379 666,width=3cm]{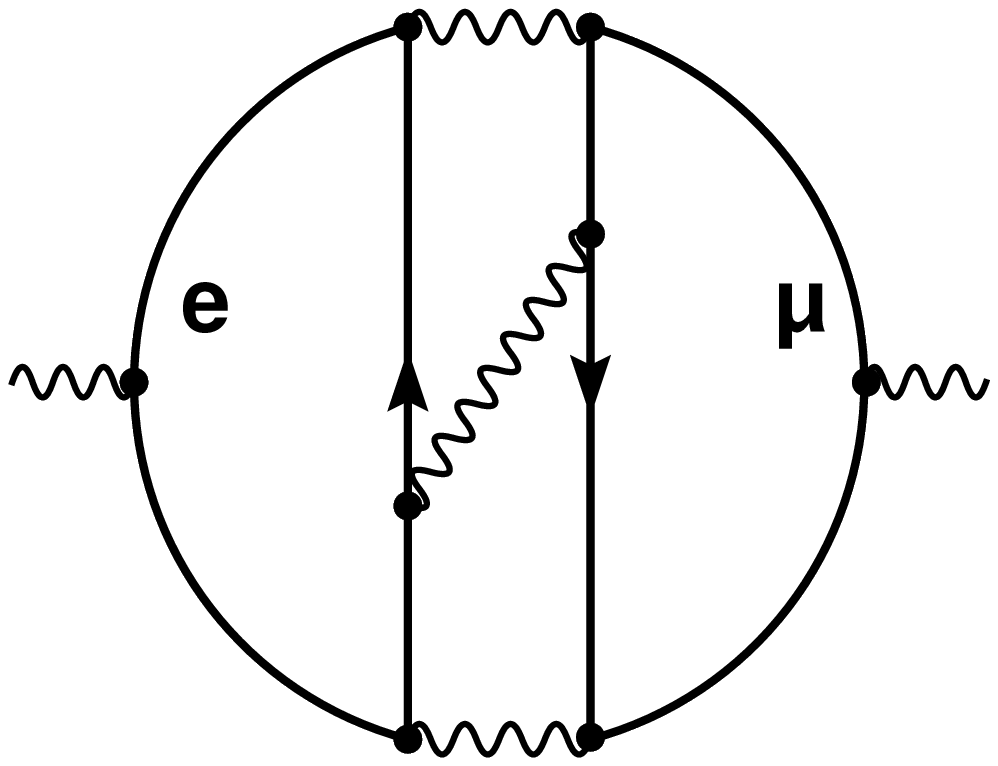}
           }\\[-0.1cm]
\hspace{2ex}$\Pi_{s,nd}^{(3)}$
\end{center}
\end{minipage}
\end{center}
\vspace{-0.4cm}
\caption{\label{fig:dia} For each of the different terms in
  Eq.~(\ref{eq:pi4loop}) is shown here one typical four-loop diagram as
  representative of the whole class.  The wavy lines denote photons; the
  lines with an arrow are leptons.  Dashed lines represent particles
  which have a smaller mass than particles which are given by solid
  lines. Dotted lines denote particles which have a smaller mass than
  particles which are represented by dashed lines. The singlet type
  diagrams ($\Pi_{s,d}^{(3)},\;\Pi_{s,nd}^{(3)}$) contribute for the
  first time at four-loop order in QED.
}
\end{figure}

The non-singlet contributions of the vacuum polarization function in the
$\MSbar$ scheme at four-loop order in QCD and at $q^2=0$ were computed
in Ref.~\cite{Chetyrkin:2006xg}.  The QED result was presented in
Refs.~\cite{Baikov:2008si,Baikov:2012rr}. Higher moments of the low
energy expansion in $q^2/\overline{m}^2$ were determined in
Refs.~\cite{Chetyrkin:2004fq,Chetyrkin:2006xg,Boughezal:2006px,Czakon:2007qi,Maier:2008he,Maier:2009fz,Kiyo:2009gb}
at four-loop order.
In these works the contributions from quenched diagrams were determined
as well as the contributions from diagrams with inserted fermion loops,
where the internal fermions were massless or had the same mass as the
external one.  In order to perform the determination of
$\Delta\alpha_{\mbox{\scriptsize{lep}}}$ in the previously described
expansion at four-loop order in QED one needs in addition still to
compute the contributions at $q^2=0$ from diagrams where the internal
fermion loops have a larger mass than the external one as well as the
non-diagonal singlet contributions. In Fig.~\ref{fig:dia} we show one
four-loop example-diagram for each arising gauge invariant subset of
diagrams. These gauge invariant subsets are given
by the different terms in Eq.~(\ref{eq:pi4loop}).\\
The two ingredients, the vacuum polarization function in the $\MSbar$
scheme in the massless limit and at $q^2=0$, are then combined to
construct the vacuum polarization function in the on-shell scheme
\begin{equation}
\label{eq:RenPi}
\Pi_{\os}(q^2,\alpha,M)=\!
{\aemOSvpi}\*
\sum_{k=0}\!\aembarpi^{k}\!\left[
               \overline{\Pi}^{(k)}(q^2,\overline{m}=0)-
               \overline{\Pi}^{(k)}(q^2=0,\overline{m})
                     \right]
            \!\bigg|_{\substack{\overline{m}=C_{\overline{m}M} M\\
              \hspace*{-1.5ex}\bar{\alpha}=C_{\bar{\alpha}\alpha}\alpha}}
\!\!\!\!+\!\mathcal{O}\left(\!{M^2\over q^2}\!\right),
\end{equation}
where the `bar' on the r.h.s. indicates that a quantity has been
renormalized in the $\MSbar$ scheme, e.g. $\bar{\alpha}$ is the
fine-structure constant in the $\MSbar$ scheme. Masses in the $\MSbar$
scheme are denoted by small letters $\overline{m}$, whereas masses in
the on-shell scheme are given by capital letters~$M$.  The conversion
function $C_{\overline{m}M}$($C_{\bar{\alpha}\alpha}$) converts the
lepton mass(fine-structure constant) from the on-shell to the $\MSbar$
scheme or vice versa.

In order to compute $\overline{\Pi}^{(3)}(q^2=0,\overline{m})$ at
four-loop order in QED we start first to generate the required diagrams
with the program {\tt{QGRAF}}~\cite{Nogueira:1991ex}. The expansion
around $q^2=0$ is performed with the programs {\tt{q2e}} and
{\tt{exp}}~\cite{Seidensticker:1999bb,Seidensticker:2001th,Harlander:1997zb}
which also map the resulting loop integrals onto a proper notation,
which allows us to perform further simplifications of the resulting loop
integrals in a straightforward way. In the next step the integrals are
reduced to a small set of master integrals with the traditional
integration-by-parts(IBP) method in combination with Laporta's
algorithm~\cite{Laporta:1996mq,Laporta:2001dd}. For the algebraic
manipulations we use
{\tt{FORM}}~\cite{Vermaseren:2000nd,Vermaseren:2002rp,Tentyukov:2006ys}
and the rational functions in the space-time dimension $d$, which arise
while solving the linear system of IBP equations, are simplified with
{\tt{FERMAT}}~\cite{Lewis:rh}. At the end of this procedure all
four-loop integrals can be expressed in terms of 12 master integrals
from which three are factorized.  They are known since long in the
literature~\cite{Schroder:2005va,Chetyrkin:2006dh}; also other authors
have contributed to individual master integrals or specific orders in
the $\vep=2-d/2$ expansion with analytic
results\cite{Broadhurst:1991fi,Schroder:2005db,Chetyrkin:2004fq,Broadhurst:1996az,Laporta:2002pg,Kniehl:2005yc,Kniehl:2006bf,Kniehl:2006bg}
which allow to obtain $\overline{\Pi}^{(3)}(q^2=0,\overline{m})$ in
Eq.~(\ref{eq:RenPi}) completely analytically.

To check our calculation we have kept the dependence on the gauge
parameter of the photon propagator and have verified its cancellation in
the final result. We have also checked the Ward-Takahashi identity and
verified that the longitudinal part of the vacuum polarization function
vanishes.

For the conversion of the vacuum polarization function to the on-shell
scheme in Eq.~(\ref{eq:RenPi}) the conversion functions
$C_{\overline{m}M}$ for the lepton mass and $C_{\bar{\alpha}\alpha}$ for
the fine-structure constant are needed.

\begin{figure}[!ht]
\begin{center}
\begin{minipage}{3cm}
\begin{center}
\includegraphics[bb=129 540 365 669,width=3cm]{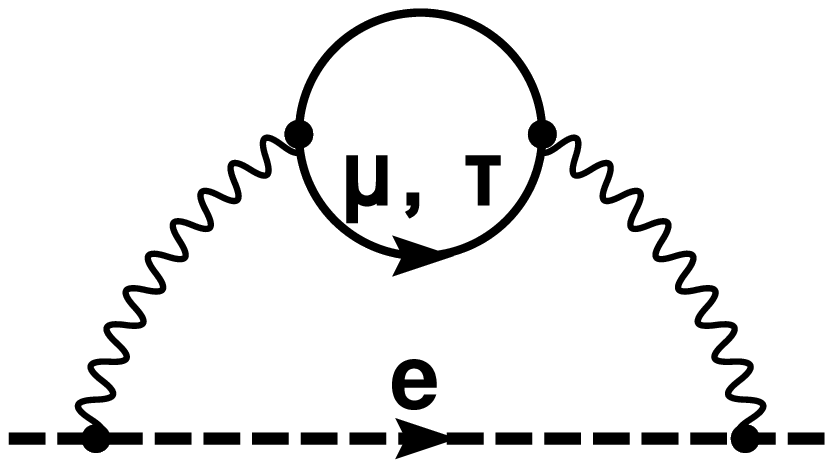}\\
(a)
\end{center}
\end{minipage}
\hspace*{1cm}
\begin{minipage}{3cm}
\begin{center}
\includegraphics[bb=129 540 365 669,width=3cm]{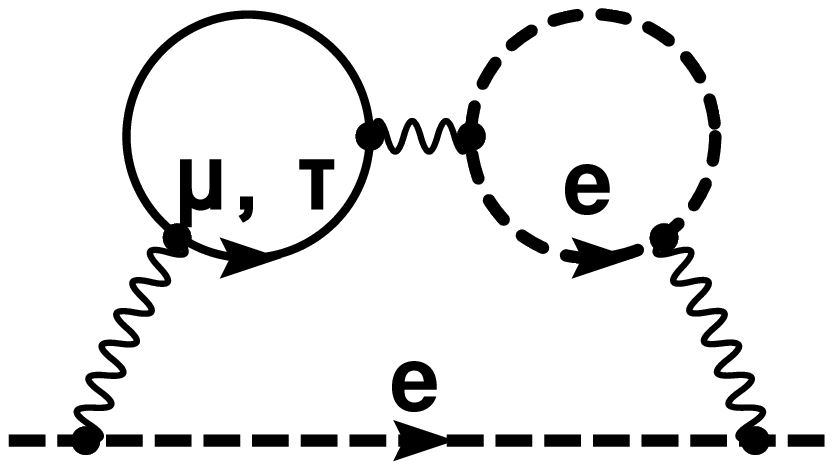}\\
(b)
\end{center}
\end{minipage}
\hspace*{0.3cm}
\begin{minipage}{3cm}
\begin{center}
\includegraphics[bb=129 540 365 669,width=3cm]{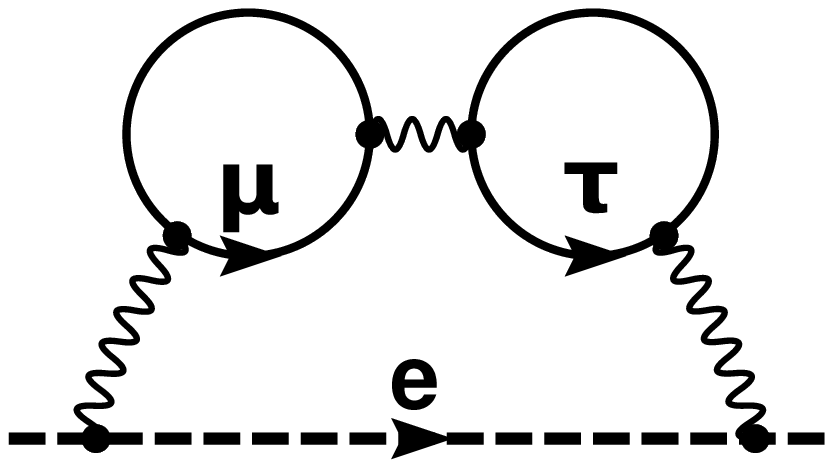}\\
(c)
\end{center}
\end{minipage}
\hspace*{0.5cm}
\end{center}
\vspace*{-0.4cm}
\caption{\label{fig:fermionself} Lepton self-energy example-diagrams at
  two-loop and three-loop order in QED which contribute to
  Eq.~(\ref{eq:MSbarOnShellMass}). The wavy lines are photons, the
  dashed and solid lines are leptons. Leptons with a dashed line have a
  smaller mass than leptons with a solid line.  }
\end{figure}

Let us start with the conversion of the lepton masses.  For the case of
the mass of the $\tau$-lepton the QED $\MSbar$--on-shell relation is
known to three-loop order; it can be deduced from the QCD results of
Refs.~\cite{Tarrach:1980up,Gray:1990yh,Chetyrkin:1999ys,Chetyrkin:1999qi,Melnikov:2000zc,Marquard:2007uj,Bekavac:2007tk}.
We decompose the different orders in perturbation theory as follows
\begin{equation}
\label{eq:MSbarOnShellMass}
\overline{m}_{\ell}=\M{\ell}\*\underbrace{\Bigg[ 1 
 + \aembarpi\*c_{\ell}^{(1)}
 + \aembarpi^2\*c_{\ell}^{(2)}
 + \aembarpi^3\*c_{\ell}^{(3)}
 + \mathcal{O}\left(\bar{\alpha}^4\right)
\Bigg]}_{=C_{\overline{m}M}}\,.
\end{equation}
The coefficient functions $c_{\ell}^{(k)}$ ($k=1,2,3$) depend in general
on the renormalization scale $\mu$ and on the lepton masses, where we
have omitted these function arguments for simplicity.  The one-loop
coefficient is known since long~\cite{Tarrach:1980up} and given in
Eq.~(\ref{eq:massconv1loop}).  Starting from two-loop order also
internal lepton loops can arise, like shown in
Fig.~\ref{fig:fermionself}. In general, the coefficient $c_{\ell}^{(2)}$
can be decomposed into gauge invariant contributions which arise from
quenched ($A$) diagrams, contributions from diagrams with light ($l$) or
heavy ($h$) inserted lepton loops and diagrams where the internal lepton
loop has the same mass as the external one ($F$).

The tauon is the heaviest lepton of the SM, so that only light internal
leptons or the tauon itself can contribute in its conversion relation of
Eq.~(\ref{eq:MSbarOnShellMass}). At two-loop order one obtains
\begin{eqnarray}
\label{eq:twoloopmasstau}
c_{\tau}^{(2)}(\M{\tau})=
      c_A^{(2)}(\M{\tau})
 + 2\*c_l^{(2)}(\M{\tau})
 +    c_F^{(2)}(\M{\tau})\,.
\end{eqnarray}
The individual terms on the r.h.s. of Eq.~(\ref{eq:twoloopmasstau}) are
known~\cite{Gray:1990yh} and are given in
Appendix~\ref{app:MSbarOnShellMass}.

At three-loop order the same decomposition with respect to the number of
inserted closed lepton loops can be made
\begin{eqnarray}
\label{eq:threeloopmasstau}
c_{\tau}^{(3)}(\M{\tau})\!&=&\!
      c_A^{(3)}(\M{\tau})
 + 2\*c_l^{(3)}(\M{\tau})
 +    c_F^{(3)}(\M{\tau})
 + 2\*c_{Fl}^{(3)}(\M{\tau})
 + 4\*c_{ll}^{(3)}(\M{\tau})
 +    c_{FF}^{(3)}(\M{\tau}),
\end{eqnarray}
where the last three terms with two letters in the subscript arise from
diagrams with two lepton loop insertions. The coefficients on the
r.h.s. of Eq.~(\ref{eq:threeloopmasstau}) are again
known~\cite{Chetyrkin:1999ys,Chetyrkin:1999qi,Melnikov:2000qh,Marquard:2007uj,Bekavac:2007tk}
and given for completeness in Appendix~\ref{app:MSbarOnShellMass}. In
particular the complete effects from a second virtual massive fermion in
the inner loops were determined at two- and three-loop order in
Refs.~\cite{Gray:1990yh,Bekavac:2007tk}. Since our final result for
$\Dalpha_{\mbox{\scriptsize{lep}}}$ at four-loop order will be in an
expansion in $M^2/M_Z^2$ up to the constant order, we do not consider
higher order virtual mass effects in Eqs.~(\ref{eq:twoloopmasstau}) and
(\ref{eq:threeloopmasstau}) as well as in the following
Eqs.~(\ref{eq:muondecoupl2loop}), (\ref{eq:cmu3}),
(\ref{eq:electrondecoupl2loop}) and (\ref{eq:ce3}).

For the conversion relation of the muon internal heavy $\tau$-lepton
loops can arise starting from two-loop order. These internal
$\tau$-lepton contributions have been determined in
Ref.~\cite{Steinhauser:1998rq} and lead to the additional contribution
$c_h^{(2)}(\M{\mu},\M{\tau})$
\begin{eqnarray}
\label{eq:muondecoupl2loop}
c_{\mu}^{(2)}(\M{\mu},\M{\tau})=
   c_A^{(2)}(\M{\mu})
 + c_l^{(2)}(\M{\mu})
 + c_F^{(2)}(\M{\mu})
 + c_h^{(2)}(\M{\mu},\M{\tau})\,.
\end{eqnarray}
Similarly also at three-loop order the new contributions with heavy
fermion loop insertions arise
\begin{eqnarray}
\label{eq:cmu3}
c_{\mu}^{(3)}(\M{\mu},\M{\tau})&=&
   c_A^{(3)}(\M{\mu})
 + c_l^{(3)}(\M{\mu})
 + c_F^{(3)}(\M{\mu})
 + c_h^{(3)}(\M{\mu},\M{\tau})
\nonumber\\
&+&c_{Fl}^{(3)}(\M{\mu})
 + c_{lh}^{(3)}(\M{\mu},\M{\tau})
 + c_{Fh}^{(3)}(\M{\mu},\M{\tau})
\nonumber\\
&+&c_{ll}^{(3)}(\M{\mu})
 + c_{FF}^{(3)}(\M{\mu})
 + c_{hh}^{(3)}(\M{\mu},\M{\tau}).
\end{eqnarray}
The coefficient functions $c_h^{(3)}(\M{\mu},\M{\tau})$,
$c_{hh}^{(3)}(\M{\mu},\M{\tau})$ and $c_{Fh}^{(3)}(\M{\mu},\M{\tau})$
will be presented in Section~\ref{sec:Results} and have been determined
with the help of the decoupling relations for the fine-structure
constant and the fermion mass
\begin{eqnarray}
\label{eq:decouplalpha}
\bar{\alpha}^{(n_f-1)}(\mu)&=&\zeta_{\gamma}^2\!\left(\mu,\bar{\alpha}^{(n_f)}(\mu),\overline{m}_h\right)\,\bar{\alpha}^{(n_f)}(\mu)\,,\\
\label{eq:decouplmass}
\overline{m}^{(n_f-1)}(\mu)&=&\zeta_{m}\!\left(\mu,\bar{\alpha}^{(n_f)}(\mu),\overline{m}_h\right)
\,\overline{m}^{(n_f)}(\mu)\,.
\end{eqnarray}
The decoupling functions
$\zeta_{\gamma}^2(\mu,\bar{\alpha}^{(n_f)}(\mu),\overline{m}_h)$ and
$\zeta_{m}(\mu,\bar{\alpha}^{(n_f)}(\mu),\overline{m}_h)$ can be
obtained from the known QCD
results~\cite{Wetzel:1981qg,Bernreuther:1981sg,Bernreuther:1983zp,Bernreuther:1983hf,Larin:1994va,Chetyrkin:1997un,Chetyrkin:2005ia,Schroder:2005hy}
and are given in Appendix~\ref{app:decouple} for completeness. The mass
$\overline{m}_h$ is the mass of the decoupled heavy fermion and the
superscript $n_f$ denotes the number of active lepton flavors. As a
check of this procedure we reproduced in Eq.~(\ref{eq:muondecoupl2loop})
the known two-loop result of Ref.~\cite{Steinhauser:1998rq} for
$c_h^{(2)}(\M{\mu},\M{\tau})$. It can also be obtained from the
results in Refs.~\cite{Gray:1990yh,Bekavac:2007tk}, which incorporate the
complete fermion mass dependence. In the same way we derived also
$c_{lh}^{(3)}(\M{\mu},\M{\tau})$ and compared it with the analytic
results of Ref.~\cite{Bekavac:2007tk}.  The remaining coefficient
functions in Eq.~(\ref{eq:cmu3}) are again known and given in
Appendix~\ref{app:MSbarOnShellMass}.

In the case of the mass conversion relation from the $\MSbar$ to the
on-shell scheme for the electron the internal loops in the Feynman
diagrams cannot have a lower mass than the external fermion, since only
internal electrons or heavy muons and $\tau$-leptons can appear. These
situations are shown in Fig.~\ref{fig:fermionself}.
\\
Similarly to Eqs.~(\ref{eq:twoloopmasstau}) and
(\ref{eq:muondecoupl2loop}) the two-loop contribution reads
\begin{eqnarray}
\label{eq:electrondecoupl2loop}
c_{e}^{(2)}(\M{e},\M{\mu},\M{\tau})&=&
   c_A^{(2)}(\M{e})
 + c_F^{(2)}(\M{e})
 + c_h^{(2)}(\M{e},\M{\mu})
 + c_h^{(2)}(\M{e},\M{\tau})
\end{eqnarray}
and the three-loop term follows in analogy to
Eqs.~(\ref{eq:threeloopmasstau}) and (\ref{eq:cmu3})
\begin{eqnarray}
\label{eq:ce3}
c_{e}^{(3)}(\M{e},\M{\mu},\M{\tau})&=&
   c_A^{(3)}(\M{e})
 + c_F^{(3)}(\M{e})
 + c_h^{(3)}(\M{e},\M{\mu})
 + c_h^{(3)}(\M{e},\M{\tau})
\nonumber\\
&+&c_{Fh}^{(3)}(\M{e},\M{\mu})
 + c_{Fh}^{(3)}(\M{e},\M{\tau})
 + c_{\mu\tau}^{(3)}(\M{e},\M{\mu},\M{\tau})
\nonumber\\
&+&c_{FF}^{(3)}(\M{e})
 + c_{hh}^{(3)}(\M{e},\M{\mu})
 + c_{hh}^{(3)}(\M{e},\M{\tau})\,.
\end{eqnarray}
The new contributions with heavy fermion loop insertions are the terms
with the coefficient functions which have a subscript $h$ as well as the
term $c_{\mu\tau}^{(3)}(\M{e},\M{\mu},\M{\tau})$.  The latter
arises from diagrams like the one shown in
Fig.~\ref{fig:fermionself}(c).  These contributions have again been
obtained with the decoupling relations of Eqs.~(\ref{eq:decouplalpha})
and (\ref{eq:decouplmass}) and are given in the next section.

We have checked that the dependence of the conversion relations on the
renormalization scale $\mu$ obeys the mass anomalous
dimension~\cite{Tarrach:1980up,Tarasov:1982aa,Chetyrkin:1997dh,Vermaseren:1997fq}.

As last ingredient the conversion factor $C_{\bar{\alpha}\alpha}$ which
is mediating between the fine-structure constant renormalized in the
$\MSbar$ and the on-shell scheme,
$\bar{\alpha}(\mu)=C_{\bar{\alpha}\alpha}\*\alpha$, is needed to order
$\alpha^2$. It is known in
literature~\cite{Broadhurst:1991fi,Baikov:2008si,Baikov:2012rr} to
four-loop order and can be obtained with the help of the concept of the
invariant charge~\cite{Bogolyubov:1956gh,Shirkov:1998ak} by computing
the vacuum polarization function at $q^2=0$ in the $\MSbar$ scheme. The
conversion factor respecting the mass hierarchy of the three charged
leptons reads
\begin{equation}
\label{eq:Cbaralphaalpha}
C_{\bar{\alpha}\alpha}=
   1
+
{\alpha\over\pi}\*\sum_{j=e,\mu,\tau}\*{\LLmu{j}\over3}
 + \left({\alpha\over\pi}\right)^2\*\sum_{j=e,\mu,\tau}\*\left[
   {15\over16} + {\LLmu{j} \over4}
 + \sum_{k=e,\mu,\tau}\*{\LLmu{j}\over3}\*{\LLmu{k}\over3}\right]
 + \mathcal{O}(\alpha^3)\,.
\end{equation}
The symbol $\LLmu{}$ denotes the logarithm $\LLmu{}=\log(\mu^2/M^2)$. 

We have checked by combining the result of the vacuum polarization
function in the $\MSbar$ scheme up to four-loop order at $q^2=0$ with
the above $\MSbar$ to on-shell conversion relations for the lepton
masses and the fine-structure constant that $\Pi_{\os}(q^2=0)$ is zero.
We have also checked that the renormalization scale dependence $\mu$ of
the vacuum polarization function cancels after the conversion to the
on-shell scheme.
\section{Results\label{sec:Results}}
In this section we summarize our results obtained with the methods
described in the previous section.  We begin with the terms due to
heavier leptons in the conversion relation of the lepton masses between
the $\MSbar$ and the on-shell scheme at three-loop order. The results
which are known in literature are listed for completeness in
Appendix~\ref{app:MSbarOnShellMass}.  The three terms which contribute
in Eqs.~(\ref{eq:cmu3}) and (\ref{eq:ce3}) read
\begin{eqnarray}
\label{eq:ch3}
c_h^{(3)}(\M{\ell},\M{x})&=&
   {547\over192}
 + {11\over360}\*\pi^4 
 + {\pi^2\over6}\*\Logn{2}{2} 
 - {1\over6}\*\Logn{2}{4} 
 - {57\over32}\*\z3 
 - 4\*\A{4}
\nonumber\\&&
 + {359\over384}\*\LLmu{\ell}
 + \LLmu{x}\*\left[ 
   {1\over48}
 + {5\over24}\*\pi^2
 - {\pi^2\over3}\*\Log{2}
\right.\nonumber\\&&\left.
 - {1\over4}\*\z3
 - {1\over16}\*\LLmu{\ell}\*\left(
   {13\over2}
 + 3\*\LLmu{\ell} 
 - {3\over2}\*\LLmu{x}
               \right)
       \right]\,,\\
\label{eq:chh3}
c_{hh}^{(3)}(\M{\ell},\M{x})&=&
   {1685\over7776}
 - {7\over18}\*\z3
 + \LLmu{x}\*\left[
   {31\over108}          
\right.\nonumber\\&&\left.
 - \LLmu{x}\*\left(
   {13\over72}           
 + {1\over12}\*\LLmu{\ell}
 - {1\over18}\*\LLmu{x}
               \right)
               \right]\,,\\
\label{eq:cFh3}
c_{Fh}^{(3)}(\M{\ell},\M{x})&=&
 - {1327\over3888} 
 + {2\over9}\*\z3
 -  \LLmu{x}\*\left[
   {5\over8}
 - {\pi^2\over9}
\right.\nonumber\\&&\left.
 + {1\over12}\*\LLmu{\ell}\*\left(
   {13\over3} + \LLmu{\ell}
              \right)
 - {1\over36}\*\LLmun{x}{2}
               \right]\,,\\
\label{eq:cmutau3}
c_{\mu\tau}^{(3)}(\M{e},\M{\mu},\M{\tau})&=&
 - {1327\over3888}
 + {2\over9}\*\z3
 + {1\over 6}\*\LLmu{\tau}\*\left[
   {31\over 9}
 + {1\over 6}\*\LLmun{\tau}{2}
\right.\nonumber\\&&\left.
 - \LLmu{\mu}\*\left(
   {13\over6}
 + \LLmu{e}
 - {1\over2}\*\LLmu{\mu}
                  \right)
                           \right]\,,
\end{eqnarray}
with $\LLmu{}=\LmudMs{}$. Here we defined
$\A{n}=\mbox{Li}_{n}(\frac{1}{2})$ with the polylogarithm function
$\mbox{Li}_{n}(z)=\sum_{k=1}^{\infty}\frac{z^k}{k^n}$. The symbol
$\zeta_n$ is the Riemann zeta function
$\z{n}=\sum_{k=1}^{\infty}k^{-n}$.  The lepton mass $\M{x}$ in
Eqs.~(\ref{eq:ch3})-(\ref{eq:cFh3}) is always larger than the lepton
mass $\M{\ell}$.

With these results one can derive the leptonic contributions to the
on-shell vacuum polarization function from all three charged lepton
generations at four-loop order in QED in an expansion for large external
momentum.  As a check of our computer routines which perform the
algebraic manipulations we reproduce the known one-loop and two-loop
results as well as the three-loop results $\Pi_{A}^{(2)}(q^2,M_i)$,
$\Pi_{l}^{(2)}(q^2,M_i,M_j)$, $\Pi_{F}^{(2)}(q^2,M_i)$,
$\Pi_{h}^{(2)}(q^2,M_i,M_j)$ which are given in Eqs.~(5)-(10) of
Ref.~\cite{Steinhauser:1998rq} and list them for completeness in
Appendix~\ref{sec:pi}. The new four-loop results in
Eq.~(\ref{eq:pi4loop}) read
\begin{eqnarray}
   \Pi_{h}^{(3)}(q^2,M_j)&=&
 - {1\over12}\*\LLn{j}{2}
 - \LL{j}\*\left(
   {1\over3} 
 + {19\over3}\*\z3 
 - {20\over3}\*\z5
                         \right) 
 + {133\over24}
\nonumber\\
&+&{5\over108}\*\pi^4 
 + {1657\over72}\*\z3 
 - 4\*\z3^2 
 - {250\over9}\*\z5
 + {2\over9}\*\pi^2\*\Logn{2}{2} 
\nonumber\\
&-&{2\over9}\*\Logn{2}{4} 
 - {16\over3}\*\A{4}
 + \dots,\\
   \Pi_{l}^{(3)}(q^2,M_i,M_j)&=&
   {1\over12}\*\LLn{i}{2}
 - {1\over6}\*\LL{j}\*\LL{i}
 + \LL{i}\*\left( 
   {41\over36} 
 + {5\over9}\*\pi^2 
 - {8\over9}\*\pi^2\*\Log{2}
\right.\nonumber\\&-&\left. 
   {29\over24}\*\z3      
          \right)
 - \LL{j}\*\left(
   {53\over36} 
 + {5\over9}\*\pi^2 
 - {8\over9}\*\pi^2\*\Log{2}
 + {41\over8}\*\z3 
 - {20\over3}\*\z5
                        \right) 
\nonumber\\
&+&{553\over108} 
 + {107\over54}\*\pi^2 
 - {919\over12960}\*\pi^4 
 + {4117\over144}\*\z3 
 - 4\*\z3^2 
 - {250\over9}\*\z5
\nonumber\\
&+&{7\over108}\*\Logn{2}{4} 
 + {89\over108}\*\pi^2\*\Logn{2}{2} 
 - {88\over27}\*\Log{2}\*\pi^2
 + {14\over9}\*\A{4}
 + \dots,\\
\label{eq:PiFh3}
   \Pi_{Fh}^{(3)}(q^2,M_i,M_j)&=&
   {1\over27}\*\LLn{j}{3}
 - {1\over9}\*\LL{i}\*\LLn{j}{2}
 + \LL{i}\*\LL{j}\*\left( {11\over9} - {8\over9}\*\z3 \right)
\nonumber\\
&-&\LL{j}\*\left( {271\over81} - {76\over27}\*\z3 \right)
 - \LL{i}\*\left( {37\over9} - {76\over27}\*\z3  \right)
 + {3956\over243} 
\nonumber\\
&-&{776\over81}\*\z3 
 - {40\over9}\*\z5
 + \dots,\\
   \Pi_{Fl}^{(3)}(q^2,M_i,M_j)&=&
   {1\over27}\*\LLn{i}{3}
 - {1\over9}\*\LL{j}\*\LLn{i}{2}
 + \LL{j}\*\LL{i}\*\left( {11\over9} - {8\over9}\*\z3 \right)
\nonumber\\
&-&\LL{i}\*\left( {703\over108} - {8\over27}\*\pi^2 - {671\over216}\*\z3 \right)
 - \LL{j}\*\left( {307\over324} + {8\over27}\*\pi^2 - {545\over216}\*\z3 \right) 
\nonumber\\
&+&{23689\over1944} 
 + {26\over81}\*\pi^2 
 - {49\over4320}\*\pi^4 
 - {2663\over324}\*\z3 
 - {40\over9}\*\z5
\nonumber\\
&-&{1\over36}\*\pi^2\*\Logn{2}{2}
 + {1\over36}\*\Logn{2}{4}
 + {2\over3}\*\A{4}
 + \dots,\\
   \Pi_{hh}^{(3)}(q^2,M_j)&=&
 - {1\over 27}\*\LLn{j}{3}
 + \LLn{j}{2}\*\left( {11\over 18} - {4\over9}\*\z3 \right)
 - \LL{j}\*\left( {302\over 81} - {76\over 27}\*\z3 \right)
\nonumber\\
&+&{4207\over 486}
 - {442\over 81}\*\z3 
 - {20\over 9}\*\z5
 + \dots,\\
   \Pi_{ll}^{(3)}(q^2,M_i,M_j)&=&
 - {1\over 27}\*\LLn{i}{3}
 + {1\over 9}\*\LL{i}\*\LL{j}\*\left(\LL{i} - \LL{j}\right)
 - {14\over 27}\*\LLn{i}{2} 
\nonumber\\
&+&\LLn{j}{2}\*\left( {5\over 54} - {4\over 9}\*\z3 \right) 
 + {28\over 27}\*\LL{j}\*\LL{i}
 - \LL{i}\*\left( {70\over 81} + {4\over 27}\*\pi^2 \right)
\nonumber\\
&-&\LL{j}\*\left( {232\over 81} - {4\over 27}\*\pi^2 - {76\over27}\*\z3 \right)
 + {3329\over 486} 
 - {26\over 81}\*\pi^2 
\nonumber\\
&-&{442\over 81}\*\z3 
 - {20\over 9}\*\z5
 + \dots,\\
\label{eq:pi3tau}
   \Pi_{e\mu}^{(3)}(q^2,M_\tau,M_e,M_\mu)&=& 
 - {2\over27}\* \LLn{\tau}{3} 
 - {2\over9}\*\LL{e}\*\LL{\mu}\*\LL{\tau} 
 + {1\over9}\*\LLn{\tau}{2}\*\left(\LL{e} + \LL{\mu}\right)
\nonumber\\
&-&{28\over27}\*\LLn{\tau}{2} 
 + {28\over27}\*\LL{\tau}\*\left(\LL{e} + \LL{\mu}\right)
 + \LL{e}\*\LL{\mu}\*\left({5\over27}-{8\over9}\*\z3\right) 
\nonumber\\
&-&\left(\LL{e}+\LL{\mu}\right)\*\left(
   {232\over81} 
 - {4\over27}\*\pi^2 
 - {76\over27}\*\z3 
          \right) 
 - \LL{\tau}\*\left(
   {140\over81} 
\right.\nonumber\\&+&\left.
  {8\over27}\*\pi^2
             \right)
 + {3329\over243} 
 - {52\over81}\*\pi^2 
 - {884\over81}\*\z3 
 - {40\over9}\*\z5
 + \dots,\\
\label{eq:pi3mue}
\Pi_{j\tau}^{(3)}(q^2,M_j,M_\tau)&=&
   {1\over 27}\*\LLn{\tau}{3}
 - {1\over9}\*\LLn{\tau}{2}\*\LL{j}
 + \LL{j}\*\LL{\tau}\*\left( {11\over9} - {8\over9}\*\z3 \right)
\nonumber\\
&-&\LL{\tau}\*\left( {271\over81} - {76\over27}\*\z3 \right)
 - \LL{j}\*\left( {37\over9} - {76\over27}\*\z3 \right)
 + {3956\over 243} 
\nonumber\\
&-&{776\over 81}\*\z3 
 - {40\over 9}\*\z5 
 +\dots,\;j=\{e,\mu\}\\
\Pi_{s,nd}^{(3)}(q^2,M)&=&
 - \LL{}\*\left({11\over9} - {8\over3}\*\z{3}\right) 
 + {82\over27}
 - {80\over9}\*\z{3} 
 - {8\over3}\*\z{3}^2 
 + {20\over3}\*\z{5}
 + \dots,
\end{eqnarray}
with the logarithm $\LL{}=\log(-q^2/M^2)$. The dots stand for a deeper
expansion in the ratio of the lepton masses and the external momentum
squared, $M^2/q^2$. The remaining symbols are the same as those defined
after Eqs.~(\ref{eq:ch3})-(\ref{eq:cmutau3}).

In the following we use Eq.~(\ref{eq:DefDeltaAlpha}) to perform a
numerical evaluation of $\Dalpha_{\mbox{\scriptsize{lep}}}$ by using the
available results of the vacuum polarization function in the on-shell
scheme up to four-loop order in order to study the relative size of the
different contributions.  For the numerical evaluation we use the
following input values from CODATA~\cite{Mohr:2012tt} and
PDG~\cite{Beringer:1900zz} for the lepton masses
$\Me=0.510998928(11)$~MeV, $\Mm=105.6583715(35)$~MeV,
$\Mt=1776.82(16)$~MeV, and for the $Z$-boson mass $M_Z=91.1876(21)$~GeV.
With these values the coefficients of the different orders in
perturbation theory of $\Dalpha_{\mbox{\scriptsize{lep}}}$ read
\begin{eqnarray}
\label{eq:evalnumaemcoeff}
\Dalpha_{\mbox{\scriptsize{lep}}}&=&
   \left(\aemOSpi\right)^{\phantom{1}} 13.52631(8)
 + \left(\aemOSpi\right)^2           14.38553(6)
\nonumber\\
&+&\left(\aemOSpi\right)^3           84.8285(7)
 + \left(\aemOSpi\right)^4 [810.65(1)_{\mbox{\scriptsize{NS}}} -39.8893(5)_{\mbox{\scriptsize{SI}}}]
 + \mathcal{O}(\alpha^5)\,,
\end{eqnarray}
where we have decomposed the new four-loop term into the contribution
from non-singlet{\scriptsize{(NS)}} and singlet{\scriptsize{(SI)}} type
of diagrams. The size of the contribution from the singlet diagrams is
much smaller than the one from non-singlet diagrams.  For the numerical
evaluations in this section we use for the one-loop contribution the
unexpanded result of Eq.~(\ref{eq:Pi0}).  At two- and three-loop order
we take the expanded results, where we include as many series
coefficients in the small lepton mass expansion until their contribution
becomes negligible compared to the numerical values given here. The same
holds for the corresponding values in Table~\ref{tab:deltaalphaNS}.
Some of the lowest expansion coefficients are shown in
Eq.~(\ref{eq:Pi1}) for the two-loop order and in
Eqs.~(\ref{eq:pia2})-(\ref{eq:pih2}) for the three-loop contribution. At
four loops we use the new result up to the constant order in the mass
expansion which was presented in this section in combination with the
known terms which are summarized in Appendix~\ref{app:Pi4loopOld}.

Evaluating also numerically the fine-structure constant
$\alpha=7.2973525698(24)\times 10^{-3}$, which is taken from
CODATA~\cite{Mohr:2012tt}, one obtains
\begin{equation}
\label{eq:dalphanum}
\Dalpha_{\mbox{\scriptsize{lep}}}
= 0.0314192...
+ 0.00007762...
+ 0.00000106...  
+ 0.00000002...\,,
\end{equation}
where each term stands for the next order in the perturbative expansion
of QED, i.e. the one-loop, two-loop, three-loop and four-loop
contribution.

The power suppressed terms (proportional to $M^2/M_Z^2$ and higher
powers) at two-loop order amount to about $\sim5\times 10^{-8}$ and are
smaller than the three-loop result up to the constant order in the small
mass expansion. The latter amounts to about $\sim10^{-6}$.  The power
suppressed terms at three-loop order amount to about $\sim4\times
10^{-10}$ and are again smaller than the four-loop result up to the
constant order, which is $\sim2\times 10^{-8}$.

A more detailed decomposition of the individual contributions which
arise from different gauge invariant subsets of diagrams is listed in
Table~\ref{tab:deltaalphaNS} for the non-singlet contributions and in
Table~\ref{tab:deltaalphaSI} for the four-loop singlet contributions.

%
\begin{table}[!ht]
\begin{center}
\hspace*{-0.32cm}
\begin{tabular}{cr|l|l|l|l}
\hline
\multicolumn{2}{c|}{\multirow{2}*{$10^7 \times \Dalpha_{\mbox{\scriptsize{lep}}}$} }&
\multicolumn{3}{c|}{outer leptons}& outer leptons\\
\multicolumn{2}{c|}{}& 
\multicolumn{1}{c|}{$e$} &\multicolumn{1}{c|}{$\mu$}&\multicolumn{1}{c|}{$\tau$}& 
{$e+\mu+\tau$}\\
\hline\hline
1-loop&                  &\p174346.6(4)     &\p91784.3(4)      &\p48061(1)        &\p314192(2)\\\hline\hline
2-loop&                  &\p379.8293(6)     &\p235.9993(6)     &\p160.341(2)      &\p776.170(3)\\\hline\hline
3-loop&quenched          &-0.1356831(2)     &\p-0.0939860(2)   &-0.077406200(3)   &-0.3070753(4)\\\cdashline{3-6}[1pt/1pt]
                                                                                   
\multicolumn{2}{r|}{\multirow{3}*{inner leptons $\left\{\!\!\!\begin{array}{r}
e\\\mu\\\tau\end{array}\!\!\!\right.$}}
                         &\p2.86735(1)      &\p2.66078(1)      &\p2.14729(2)      &\p7.67542(4)\\
                        &&\p0.839626(6)     &\p0.844870(6)     &\p0.823473(6)     &\p2.50797(2)\\
                        &&\p0.24972(1)      &\p0.24972(1)      &\p0.25560(1)      &\p0.75503(4)\\\hdashline
                     &sum&\p3.82101(2)      &\p3.66138(2)      &\p3.14895(4)      &\p10.63134(9)\\\hline\hline
4-loop&quenched          
  &-0.00069
  &-0.00013
  &\p0.00016
  &-0.00066%
\\\cdashline{3-6}[1pt/1pt]
\multicolumn{2}{r|}{\multirow{9}*{inner leptons $\left\{\!\!\!\begin{array}{r}
e\\\mu\\\tau\\ee\\\mu\mu\\\tau\tau\\e\mu\\e\tau\\\mu\tau\end{array}\!\!\!\right.$}}
  &\p0.00512
  &\p0.00391
  &\p0.00261
  &\p0.01164\\
 &&\p0.00197
  &\p0.00188
  &\p0.00150
  &\p0.00535\\
 &&\p0.00081
  &\p0.00081
  &\p0.00072
  &\p0.00234
\\\cdashline{3-6}[1pt/1pt]
 &&\p0.03361
  &\p0.03318
  &\p0.02955
  &\p0.09634\\
 &&\p0.00497
  &\p0.00498
  &\p0.00509
  &\p0.01504\\
 &&\p0.00061
  &\p0.00061
  &\p0.00062
  &\p0.00183%
\\\cdashline{3-6}[1pt/1pt]
 &&\p0.02376
  &\p0.02386
  &\p0.02375
  &\p0.07137\\
 &&\p0.00748
  &\p0.00748
  &\p0.00763
  &\p0.02259\\
 &&\p0.00336
  &\p0.00336
  &\p0.00342
  &\p0.01014%
\\\hdashline
&sum
 &\p0.08100
 &\p0.07994
 &\p0.07505
 &\p0.23599%
\\\hline\hline
\multicolumn{2}{c|}{1+2+3+4-loop} 
                         &\p174730.4(4)     &\p92024.0(4)       &\p48225(1)       &\p314979(2)\\\hline\hline
\end{tabular}
\caption{\label{tab:deltaalphaNS} The table shows the numerical
  evaluation of the different gauge invariant subsets which contribute
  to $\Dalpha_{\mbox{\scriptsize{lep}}}$ from one-loop to four-loop
  order. Some example-diagrams which correspond to the different
  contributions are given in Figs.~\ref{fig:12loop}, \ref{fig:3loop} and
  \ref{fig:dia}. Starting from three-loop order the table shows the
  different contributions which arise from diagrams with external(outer)
  leptons and internal(inner) leptons separately. At four-loop order
  arise for the first time diagrams with two internal lepton
  insertions. The non-singlet contributions at four-loop order are not
  given here, but are shown in Table~\ref{tab:deltaalphaSI}.}
\end{center}
\end{table}
%
At three-loop and also at four-loop order the dominant contribution to
$\Dalpha_{\mbox{\scriptsize{lep}}}$ emerges from diagrams with internal
lepton loop insertions, especially from the insertions of electrons,
which can be seen in Table~\ref{tab:deltaalphaNS}. The reason for this
dominance is an logarithmic enhancement of these terms which arises from
large logarithms in the small mass ratio $\Me^2/M_Z^2$. In comparison
the contributions from quenched diagrams are small. This behavior was
already observed at three-loop order in Ref.~\cite{Steinhauser:1998rq}
and also holds at four-loop order.
%
\begin{table}[!ht]
\begin{center}
\begin{tabular}{cr|l|l|l}
\hline
\multicolumn{2}{c|}{\multirow{2}*{$10^7 \times \Dalpha_{\mbox{\scriptsize{lep}}}$}}&
\multicolumn{3}{c}{leptons in right loop}\\
&&\multicolumn{1}{c|}{$e$}&
 \multicolumn{1}{c|}{$\mu$}&
 \multicolumn{1}{c}{$\tau$}\\
\hline\hline
\multicolumn{2}{r|}{\multirow{3}*{
\begin{minipage}{1.75cm}
leptons in \\
left loop
\end{minipage}
$\left\{\!\!\!\begin{array}{r}e\\\mu\\\tau\end{array}\!\!\!\right.$}}
  &-0.0030194
  &-0.0016177
  &-0.0008029
\\
 &&-0.0016177
  &-0.0014803
  &-0.0008029
\\
 &&-0.0008029
  &-0.0008029
  &-0.0006656
\\\hline\hline
\end{tabular}
\caption{\label{tab:deltaalphaSI}
Numerical evaluation of the non-singlet terms at four-loop order
in QED. The last two diagrams of Fig.~\ref{fig:dia} are typical 
examples which contribute to this table.}
\end{center}
\end{table}\\
\noindent
Summing up all contributions one obtains
$\Dalpha_{\mbox{\scriptsize{lep}}}=0.0314979$ with an uncertainty of
about $\sim2\cdot10^{-7}$, mainly due to the error in the $\tau$-lepton
mass and the $Z$-boson mass. This uncertainty is about one order of
magnitude bigger than the size of the new four-loop contribution, which
is about $\sim2\cdot10^{-8}$, so that the theory uncertainty in
$\Dalpha_{\mbox{\scriptsize{lep}}}$ due to the truncation of the
perturbative series has been completely removed.  The analytic formulae
which were presented in this work can be easily used to reevaluate
$\Dalpha_{\mbox{\scriptsize{lep}}}$, once more precise values of the input
parameters become available.

\section{Summary and conclusion\label{sec:DiscussConclude}}
We have determined the four-loop QED corrections to the vacuum
polarization function in the on-shell scheme in the high energy
expansion taking into account the mass hierarchy of all charged leptons
in the Standard Model. In order to derive this on-shell result we
combine known $\MSbar$ results for the vacuum polarization function in
the massless limit with newly computed contributions at $q^2=0$, which
allow to do the conversion from the $\MSbar$ to the on-shell scheme.
The on-shell result is then used to perform a determination of the
leptonic contributions to $\Dalpha$ which describes the running of the
effective electromagnetic coupling to the $Z$-boson mass scale.  The
quantity $\Dalpha_{\mbox{\scriptsize{lep}}}$ is an important ingredient
for constraining the Standard Model in the context of electroweak
precision measurements. We perform an numerical evaluation of
$\Dalpha_{\mbox{\scriptsize{lep}}}$ in order to study the relative size of the
different contributions.  The size of the new four-loop contribution is
small and completely removes the theory uncertainty due to the
truncation of the perturbative expansion. The QED corrections to
$\Dalpha_{\mbox{\scriptsize{lep}}}$ are thus from theory side well under
control compared to the error in $\Dalpha_{\mbox{\scriptsize{lep}}}$
arising from the uncertainty in the mass of the $\tau$-lepton and the
$Z$-boson mass.

As a by-product of the calculation we also determine the contributions
due to heavy fermions to the conversion relation between the $\MSbar$
and the on-shell scheme for the lepton masses in QED.

\vspace{2ex}
\noindent
{\bf Acknowledgment:}\\
C.S. would like to thank Johann K{\"u}hn for inspiring discussions,
advice and comments on the manuscript as well as Matthias
Steinhauser for communications on Ref.~\cite{Steinhauser:1998rq}
and the use of decoupling relations.\\

\noindent
The Feynman diagrams were drawn with the help of
{\tt{Axodraw}}\cite{Vermaseren:1994je} and
{\tt{Jaxodraw}}\cite{Binosi:2003yf}.
\begin{appendix}
  \section{The on-shell vacuum polarization function in the high-energy
    limit at three-loop order and known four-loop results\label{sec:pi}}
In this appendix we provide known results for the vacuum polarization
function at three-loop~\cite{Steinhauser:1998rq} and
four-loop~\cite{Baikov:2008si,Baikov:2012rr} order from literature in
order to have all components in a uniform notation.
\subsection{Three-loop results\label{app:Pi3loop}}
The results of Ref.~\cite{Steinhauser:1998rq} are given here for
completeness and read
\begin{eqnarray}
\label{eq:pia2}
\Pi_{A}^{(2)}(q^2,M_i)&=&
   {1\over8}\*\LL{i}
 - {121\over48}
 - {\pi^2\over3}\*\left( {5\over2} - 4\*\Log{2}\right) 
 - {99\over16}\*\z{3} 
 + 10\*\z{5}
 + {M_i^2\over q^2}\*\bigg[
   9\*\LL{i}^2
\nonumber\\&&
 - {3\over2}\*\LL{i}
 + {139\over 3} 
 - \pi^2\*\left( 5 - 8\*\Log{2} \right)
 - {41\over 3}\*\z3 
 - {70\over 3}\*\z5 
                    \bigg]
\nonumber\\&&
 + \left({M_i^2\over q^2}\right)^{\!2}\*\!\bigg[
  12\*\LL{i}^3
+ {27\over2}\*\LL{i}^2
+ \LL{i}\*\left({115\over4} - 2\*\pi^2\*\left(5 - 8\*\Log{2}\right) 
\right.\nonumber\\&&\left.
                 - 48\*\z3\right) 
+ {437\over6} 
- {10\over9}\*\pi^4 
- {16\over3}\*\pi^2\*\Logn{2}{2}
+ {16\over3}\*\Logn{2}{4}
+ 128\*\A{4} 
\nonumber\\&&
+ 32\*\z3 
- 20\*\z5
                                  \bigg]
 + \dots,\\
\label{eq:pif2}
\Pi_{F}^{(2)}(q^2,M_i)&=&\!\!\!
 - {1\over6}\*\LLn{i}{2}
 + \LL{i}\*\left({11\over6} - {4\over3}\*\z{3}\right)
 - {307\over216}
 - {4\over9}\*\pi^2 
 + {545\over144}\*\z{3}
 - {M_i^2\over q^2}\*\bigg[
   2\*\LL{i}^2
\nonumber\\&&
 - {26\over 3}\*\LL{i} 
 + {40\over 3} 
 + {8\over 3}\*\pi^2 
 - 16\*\z3 
                    \bigg]
 - \left({M_i^2\over q^2}\right)^{\!2}\*\!\bigg[
  {4\over3}\*\LL{i}^3 
 - {19\over3}\*\LL{i}^2 
\nonumber\\&&
 + \LL{i}\*\left( {8\over9} + {16\over3}\*\pi^2 - {40\over3}\*\z3 \right) 
 + {1505\over54}
 - {352\over9}\*\z3
                                       \bigg]
 + \dots,\\ 
\label{eq:pil2}
\Pi_{l}^{(2)}(q^2,M_i,M_j)&=&
   {1\over6}\*\LLn{i}{2}
 + {14\over9}\*\LL{i} 
 + \LL{j}\*\left( {5\over18} - {4\over3}\*\z3 \right) 
 - {1\over3}\*\LL{j}\*\LL{i}
 - {116\over27} 
\nonumber\\&&
 + {2\over9}\*\pi^2 
 + {38\over9}\*\z3
 + 2\*{M_{i}^2\over q^2}\*\left[
   \LL{i}^2
 - 2\*\LL{i}\*\LL{j}
 + {13\over3}\*\LL{i}
 - 2 
 + {2\over3}\*\pi^2 
                     \right]
\nonumber\\&&
 - 16\*{M_{j}^2\over q^2}\*\left[
   {4\over3} 
 - \z3
                     \right]
 - \left({M_{i}^2\over q^2}\right)^{\!2}\*\!\left[
   {83\over54} 
 + {224\over9}\*\z3 
 - \LL{i}\*\left({68\over9} + {8\over3}\*\pi^2\right)
 \right.\nonumber\\&&\left.
 - \LL{i}^2\*{32\over3}
 - \LL{i}^3\*{8\over3}
 - \LL{j}\*\left( {2\over9} + {16\over3}\*\z3 - {10\over3}\*\LL{i} - 4\*\LL{i}^2\right)
                                       \right]
\nonumber\\&&
 + 12\*{M_{i}^2\over q^2}\*{M_{j}^2\over q^2}\!\*\!\left[
   3 
 - 2\*\LL{i}
                     \right]
 - \left({M_{j}^2\over q^2}\right)^{\!2}\*\!\left[
   {50\over9} 
 - 16\*\z3 
 - \LL{j}^2 
\right.\nonumber\\&&\left.
 - \LL{j}\*\left(8\*\z3 - 4\*\LL{i}\right) 
 + {26\over3}\*\LL{i} 
 - 2\*\LL{i}^2
                     \right]
 + \dots,\\
\label{eq:pih2}
\Pi_{h}^{(2)}(q^2,M_i,M_j)&=&
 - {1\over6}\*\LLn{j}{2}
 + \LL{j}\*\left( {11\over6} - {4\over3}\*\z3\right)
 - {37\over 6} 
 + {38\over 9}\*\z3
 - 16\*{M_{j}^2\over q^2}\*\left[
   {4\over3}
 - \z3
                     \right]
\nonumber\\&&
 - {M_{i}^2\over q^2}\*\left[
   2\*\LL{j}^2
 - {26\over3}\*\LL{j}
 + {187\over9}
\right]
 - \left({M_{j}^2\over q^2}\right)^{\!2}\*\!\left[
   {67\over3} 
 - 16\*\z3 
\right.\nonumber\\&&\left.
 + \LL{j}\*\left({26\over3} - 8\*\z3\right)
 + \LL{j}^2
\right]
 + 12\*{M_{i}^2\over q^2}\*{M_{j}^2\over q^2}\*\left[
   1 
 - 2\*\LL{j}
\right]
\nonumber\\&&
 + \left({M_{i}^2\over q^2}\right)^{\!2}\*\!\left[
  {847\over18} 
- {224\over9}\*\z3 
- \LL{i}\*\left({10\over9} - {2\over3}\*\LL{j}\right) 
\right.\nonumber\\&&\left.
- \LL{j}\*\left({74\over3} - {16\over3}\*\z3\right) 
+ {20\over3}\*\LL{j}^2 
- {4\over3}\*\LL{j}^3
\right]
 + \dots.
\end{eqnarray}
For the contributions $\Pi_{A}^{(2)}(q^2,M_i)$ and
$\Pi_{F}^{(2)}(q^2,M_i)$ in Eqs.~(\ref{eq:pia2}) and (\ref{eq:pif2}) we
have taken the sub-leading terms in the small mass expansion from
Refs.~\cite{Chetyrkin:1994ex,Chetyrkin:1995ii,Chetyrkin:1996cf,Chetyrkin:1997qi},
which we have converted in addition from the $\MSbar$ to the on-shell
scheme.  Using {\tt{q2e}} and
{\tt{exp}}~\cite{Seidensticker:1999bb,Seidensticker:2001th,Harlander:1997zb}
we have determined also the sub-leading terms for
$\Pi_{l}^{(2)}(q^2,M_i,M_j)$ and $\Pi_{h}^{(2)}(q^2,M_i,M_j)$ in
Eqs.~(\ref{eq:pil2}) and (\ref{eq:pih2}).
The arising integrals have been computed in two ways: on the one hand
they were determined by applying Laporta's
algorithm~\cite{Laporta:1996mq,Laporta:2001dd} in order to reduce all
integrals to a set of known master integrals, and on the other hand they
were calculated by using the {\tt{MATAD}} and
{\tt{MINCER}}~\cite{Steinhauser:2000ry,Gorishnii:1989gt,Larin:1991fz}
routines.
The dots in the equations stand for terms of higher order in the small
mass expansion.  The mass $M_{i}$ in Eqs.~(\ref{eq:pil2}) and
(\ref{eq:pih2}) is the mass of the lepton in the outer loop which
connects to the external photons, whereas the mass $M_{j}$ is the mass
of the internal lepton loop. The definition of the remaining symbols
which appear in Eqs.~(\ref{eq:pia2})-(\ref{eq:pih2}) can be found in
Section~\ref{sec:Results}.
\subsection{Four-loop results\label{app:Pi4loopOld}}
The results for the vacuum polarization function of
Refs.~\cite{Baikov:2008si,Baikov:2012rr} are given here for completeness
and read
\begin{eqnarray}
   \Pi_{A}^{(3)}(q^2,M_i)&=&
   {23\over32}\*\LL{i}
 - {71189\over8640 }
 - {157\over18}\*\pi^2 
 - {59801\over32400}\*\pi^4 
\nonumber\\
&+&\pi^2\*\Log{2}\*\left({59\over3} 
 + {424\over675}\*\pi^2\right)
 - {1559\over270}\*\pi^2\*\Logn{2}{2} 
 + {128\over135}\*\pi^2\*\Logn{2}{3} 
\nonumber\\
&+&{1559\over270}\*\Logn{2}{4} 
 - {128\over225}\*\Logn{2}{5} 
 + \left({6559\over80} - {\pi^2\over6}\right)\*\z{3} 
\nonumber\\
&-&{1603\over30}\*\z{5}
 - 35\*\z{7}
 + {6236\over45}\*\A{4}
 + {1024\over15}\*\A{5}
 + \mathcal{O}\!\left(\tfrac{M_i^2}{q^2}\right),\\
\Pi_{F}^{(3)}(q^2,M_i)&=&
 - {1\over12}\*\LLn{i}{2}
 - \LL{i}\*\left({1\over3}+{19\over3}\*\z{3}-{20\over3}\*\z{5}\right)
 + {3361\over225}
 - {179\over81}\*\pi^2 
\nonumber\\
&-&{2161\over2700}\*\pi^4 
 + {64\over27}\*\pi^2\*\Log{2}
 - {53\over15}\*\pi^2\*\Logn{2}{2} 
 + {53\over15}\*\Logn{2}{4} 
\nonumber\\
&+&{29129\over450}\*\z{3} 
 - 4\*\z{3}^2 
 - {250\over9}\*\z{5} 
 + {424\over5}\*\A{4}
 + \mathcal{O}\!\left(\tfrac{M_i^2}{q^2}\right),\\
\Pi_{FF}^{(3)}(q^2,M_i)&=&
 - {1\over27}\*\LLn{i}{3}
 + \LLn{i}{2}\*\left({11\over18} - {4\over9}\*\z{3}\right)
 - \LL{i}\*\left({302\over81} - {76\over27}\*\z{3}\right)
\nonumber\\
&+&{75259\over17010} 
 + {32\over405}\*\pi^2
 - {15109\over5670}\*\z{3}
 - {20\over9}\*\z{5}
 + \mathcal{O}\!\left(\tfrac{M_i^2}{q^2}\right),\\
\Pi_{s,d}^{(3)}(q^2,M_i)&=&
 - \LL{i}\*\left({11\over9} - {8\over3}\*\z{3}\right)
 + {1963\over945}
 - {2237\over4320}\*\pi^4 
 - {73\over36}\*\pi^2\*\Logn{2}{2} 
\nonumber\\
&+&{73\over36}\*\Logn{2}{4} 
 + {5309\over280}\*\z{3}
 - {8\over3}\*\z{3}^2 
 + {20\over3}\*\z{5}
 + {146\over3}\*\A{4}
 + \mathcal{O}\!\left(\tfrac{M_i^2}{q^2}\right).
\end{eqnarray}
The definition of the different symbols can be found again in
Section~\ref{sec:Results}.
\section{Decoupling functions\label{app:decouple}}
The decoupling functions in Eqs.~(\ref{eq:decouplalpha}) and
(\ref{eq:decouplmass}) are known in QCD to high order in perturbation
theory~\cite{Wetzel:1981qg,Bernreuther:1981sg,Bernreuther:1983zp,Bernreuther:1983hf,Larin:1994va,Chetyrkin:1997un,Chetyrkin:2005ia,Schroder:2005hy}
and can be used for the QED calculation considered here after inserting
the proper values for the color factors. We give the results of the
perturbative expansion up to the order needed within this work
\begin{eqnarray}
\label{eq:zetagamma}
\zeta_{\gamma}^2\left(\mu,\bar{\alpha}^{(n_f)}(\mu),\overline{m}_h\right)&=&
   1 
 - \aemnfpi\*{\nh\over3}\*\LLmumh
 - \aemnfpi^2\!\!\*\nh\*\left(   
   {13\over48} 
\right.\nonumber\\&&\left.
 - {1\over4}\*\LLmumh 
 - {\nh\over9}\*\LLmumhn{2} 
              \right)
 + \mathcal{O}\left(\!\left({\bar{\alpha}^{(n_f)}}\right)^3\right),\\
\label{eq:zetam}
\zeta_{m}\left(\mu,\bar{\alpha}^{(n_f)}(\mu),\overline{m}_h\right)&=&
   1
 + \aemnfpi^2\*{\nh\over8}\*\left(
   {89\over36}
 - {5\over3}\*\LLmumh
 + \LLmumhn{2} 
                   \right)
\nonumber\\&&\;\;\,
 - \aemnfpi^3\!\!\*\nh\*\left[
   {683\over 576} 
 - {57\over 32}\*\z3
 + {11\over 360}\*\pi^4 
 + {\pi^2\over6}\*\Logn{2}{2} 
\right.\nonumber\\&&\;\;\,\left.
 - {1\over6}\*\Logn{2}{4}
 - 4\*\A{4}
 + \LLmumh\*\left(
   {13\over64}
 - {3\over4}\*\z3
            \right)
 + {1\over4}\*\LLmumhn{2}
\right.\nonumber\\&&\;\;\,\left.
 + {\nh\over18}\*\left(
   {1685\over432}
 - 7\*\z3
 + {31\over6}\*\LLmumh
 - {5\over4}\*\LLmumhn{2}
 + \LLmumhn{3}
                \right)
\right.\nonumber\\&&\;\;\,\left.
 - {\nl\over9}\*\left(
   {1327\over432 }
 - 2\*\z3
 - {53\over16}\*\LLmumh
 - {1\over4}\*\LLmumhn{3}
               \right) 
                  \!\right]
\!\!+\!\mathcal{O}\!\left(\!\left({\bar{\alpha}^{(n_f)}}\right)^4\right)\!\!,\;
\end{eqnarray}
with $\LLmumh=\Lmudmhs$, where $\overline{m}_h$ is the heavy particle
mass renormalized in the $\MSbar$ scheme which has been integrated out
in the effective field theory. The number of active fermions is
$n_f=\nl+\nh$ with $\nh=1$ and $\nl$ is the number of light fermions,
which are considered as massless.  The definition of the remaining
symbols in Eqs.~(\ref{eq:zetagamma}) and (\ref{eq:zetam}) can be found
in Section~\ref{sec:Calculation} and \ref{sec:Results}.
\section{The relation between the $\MSbar$ and on-shell scheme for
  lepton masses\label{app:MSbarOnShellMass}}
The known relations up to three-loop order in QED for the conversion of
the lepton masses from the $\MSbar$ to the on-shell
scheme~\cite{Tarrach:1980up,Gray:1990yh,Chetyrkin:1999ys,Chetyrkin:1999qi,Melnikov:2000zc,Marquard:2007uj,Bekavac:2007tk}
are given here for completeness. The expansion coefficient in
Eq.~(\ref{eq:MSbarOnShellMass}) at one-loop order reads
\begin{eqnarray}
\label{eq:massconv1loop}
c_{\ell}^{(1)}(\M{\ell})  &=&  - 1 - {3\over4}\*\LLmu{\ell}\,.
\end{eqnarray}
The coefficients in Eqs.~(\ref{eq:twoloopmasstau}),
(\ref{eq:muondecoupl2loop}) and (\ref{eq:electrondecoupl2loop}) at
two-loop order are given by
\begin{eqnarray}
c_A^{(2)}(\M{\ell})&=&
   {7\over 128}
 - {\pi^2\over2}\*\left({5\over 8} - \Log{2}\right)
 - {3\over 4}\*\z3 
 + {3\over 32}\*\LLmu{\ell}\*\left( 7 + 3\*\LLmu{\ell} \right)\,,\\
c_l^{(2)}(\M{\ell})&=&   
   {71\over 96 }
 + {\pi^2\over 12}
 + {13\over 24}\*\LLmu{\ell} 
 + {1\over 8}\*\LLmun{\ell}{2}\,,\\
c_F^{(2)}(\M{\ell})&=&   
   {143\over 96 }
 - {\pi^2\over6}
 + {13\over 24}\*\LLmu{\ell}
 + {1\over8}\*\LLmun{\ell}{2}\,,\\
c_h^{(2)}(\M{\ell},\M{x})&=&
 - {89\over288}
 + {13\over24}\*\LLmu{x}
 - {1\over8}\*\LLmun{x}{2}
 + {1\over4}\*\LLmu{\ell}\*\LLmu{x}\,,
\end{eqnarray}
and at three-loop order the coefficients of
Eqs.~(\ref{eq:threeloopmasstau}), (\ref{eq:cmu3}) and (\ref{eq:ce3})
read
\begin{eqnarray}
c_A^{(3)}(\M{\ell})&=&
 - {2969\over 768}
 - {613\over192}\*\pi^2 
 - {\pi^4\over48}
 + {29\over4}\*\pi^2\*\Log{2}
 + {1\over2}\*\Logn{2}{2}\*\pi^2 
\nonumber\\
&-&{1\over2}\*\Logn{2}{4} 
 - {81\over16}\*\z3 
 - {\pi^2\over16}\*\z3
 + {5\over8}\*\z5
 - 12\*\A{4}
\nonumber\\
&-&\LLmu{\ell}\*\left(
   {489\over512} 
 - {15\over64}\*\pi^2 
 + {3\over8}\*\Log{2}\*\pi^2 
 - {9\over16}\*\z3
               \right) 
\nonumber\\
&-&{27\over128}\*\LLmun{\ell}{2} 
 - {9\over128}\*\LLmun{\ell}{3}\,,\\
c_l^{(3)}(\M{\ell})&=&
   {1283\over576}
 + {13\over18}\*\pi^2
 - {119\over2160}\*\pi^4
 - {11\over9}\*\Log{2}\*\pi^2
 + {2\over9}\*\Logn{2}{2}\*\pi^2
\nonumber\\&&
 + {1\over9}\*\Logn{2}{4}
 + {55\over24}\*\z3
 + {8\over3}\*\A{4}
 + \LLmu{\ell}\*\left(
   {65\over384}
 + {7\over48}\*\pi^2  
\right.\nonumber\\&&\left.
 - {\pi^2\over3}\*\Log{2}
 - {1\over4}\*\z3\right) 
 - {13\over32}\*\LLmun{\ell}{2}
 - {3\over32}\*\LLmun{\ell}{3}\,,\\
c_F^{(3)}(\M{\ell})&=&
   {1067\over576}
 - {85\over108}\*\pi^2
 + {91\over2160}\*\pi^4 
 + {8\over9}\*\Log{2}\*\pi^2 
 - {\pi^2\over9}\*\Logn{2}{2}
\nonumber\\&&
 + {1\over9}\*\Logn{2}{4} 
 - {53\over24}\*\z3
 + {8\over3}\*\A{4}
 - {13\over32}\*\LLmun{\ell}{2}
 - {3\over32}\*\LLmun{\ell}{3}
\nonumber\\&&
 - \LLmu{\ell}\*\left(
   {151\over384}
 - {\pi^2\over3}\*\left(1 - \Log{2}\right)
 + {1\over4}\*\z3
               \right)\,,\\
c_{Fl}^{(3)}(\M{\ell})&=&
 - {5917\over3888}
 + {13\over108}\*\pi^2 
 + {2\over9}\*\z3
 - {1\over18}\*\LLmu{\ell}\*\left[
   {143\over6} 
 - \pi^2
\right.\nonumber\\&&\left.
 + \LLmu{\ell}\*\left( {13\over2} + \LLmu{\ell}\right)
            \right]\,,\\
c_{ll}^{(3)}(\M{\ell})&=&
 - {2353\over7776}
 - {13\over108}\*\pi^2
 - {7\over18}\*\z3
 - \LLmu{\ell}\*\left[
   {89\over216} 
 + {\pi^2\over18}
\right.\nonumber\\&&\left.
 + {1\over36}\*\LLmu{\ell}\*\left( {13\over2} 
                               + \LLmu{\ell} \right)
               \right]\,,\\
c_{FF}^{(3)}(\M{\ell})&=&
 - {9481\over7776}
 + {4\over135}\*\pi^2 
 + {11\over18}\*\z3
 - \LLmu{\ell}\*\left[
   {197\over216}  
 - {1\over9}\*\pi^2
\right.\nonumber\\&&\left.
 + {1\over36}\*\LLmu{\ell}\*\left( {13\over2} + \LLmu{\ell} \right)
              \right]\,,\\
c_{lh}^{(3)}(\M{\ell},\M{x})&=&
 - {1327\over3888}
 + {2\over9}\*\z3
 - \LLmu{x}\*\left[
   {1\over8} + {\pi^2\over18}
\right.\nonumber\\&&\left.
 + {1\over12}\*\LLmu{\ell}\*\left( {13\over3} + \LLmu{\ell}\right)
 - {1\over36}\*\LLmun{x}{2} 
            \right]\,,
\end{eqnarray}
with $\LLmu{}=\LmudMs{}$ and $\M{x}>\M{\ell}$. The definition of the
other symbols can be found in Section~\ref{sec:Results}. The remaining
three-loop coefficients which are needed for Eqs.~(\ref{eq:cmu3}) and
(\ref{eq:ce3}) are given in Eqs.~(\ref{eq:ch3}) to (\ref{eq:cmutau3}) of
Section~\ref{sec:Results}.  The complete mass dependence of the
contributions from diagrams with massive lepton loop insertions at two-
and three-loop order has been determined in
Refs.~\cite{Gray:1990yh,Bekavac:2007tk}. We restrict ourselves in this
section to the contributions needed for this work.
%
%
%

\providecommand{\href}[2]{#2}\begingroup\raggedright\endgroup
\end{appendix}

\end{document}